\documentclass[11pt,a4paper]{article}
\pdfoutput=1
\usepackage{mystyle}
\usepackage{amsfonts,amssymb,amsmath}
\usepackage{graphicx,epstopdf} 
\usepackage{array} 
\usepackage{placeins}
\usepackage{slashed}
\usepackage{youngtab}
\usepackage{physics}

\usepackage{IEEEtrantools}

\usepackage{hyperref}
\usepackage[color,final]{showkeys}  

\newcommand{\be}{\begin{equation}}
\newcommand{\ee}{\end{equation}}

\newcommand{\A}{\alpha}
\newcommand{\B}{\beta}

\newcommand{\lm}{\lambda}

\newcommand{\half}{\frac{1}{2}}

\newcommand{\la}{\langle}
\newcommand{\ra}{\rangle}

\newcommand{\al}{\alpha}
\newcommand{\ga}{\gamma}         \newcommand{\Ga}{\Gamma}
        
\newcommand{\ep}{\epsilon}

\def\th{\theta}       

\newcommand{\ka}{\kappa}
\newcommand{\na}{\nabla}
\newcommand{\lam}{\lambda}       
\newcommand{\bph}{\bar \phi}         \newcommand{\bPh}{\bar \Phi}
\newcommand{\bps}{\bar \psi}

\newcommand{\mN}{\mathcal N}
\newcommand{\mS}{\mathcal S}
\newcommand{\mD}{\mathcal D}
\newcommand{\mJ}{\mathcal J}
\newcommand{\nn}{\nonumber}
\newcommand{\del}{\partial}

\begin{document}

\title{Correlation functions in ${\cal N}=2$ Supersymmetric vector matter Chern-Simons  theory}
\author[a,b]{Karthik Inbasekar,}
\author[c]{Sachin Jain,}
\author[c]{Vinay Malvimat,}
\author[c]{Abhishek Mehta,}
\author[d]{Pranjal Nayak,}
\author[e,f,g]{Tarun Sharma}

\affiliation[a]{Faculty of Exact Sciences, School of Physics and Astronomy, Tel Aviv University, Ramat Aviv 69978, Israel}
\affiliation[b]{Department of Physics, Ben-Gurion University of the Negev, Beer-Sheva 84105, Israel.}
\affiliation[c]{Indian Institute of Science Education and Research, Homi Bhabha Rd, Pashan, Pune 411 008, India}
\affiliation[d]{Department of Physics \& Astronomy, Chemistry-Physics Building, 505 Rose St., University of Kentucky, Lexington, 40506, USA}
\affiliation[e]{Department of Physics, Brown University, 182 Hope Street, Providence, RI 02912, USA}
\affiliation[f]{National Institute for Theoretical Physics, School of Physics and Mandelstam Institute for Theoretical Physics, University of the Witwatersrand, Johannesburg Wits 2050, South Africa}
\affiliation[g]{School of Physical Science, National Institute of Science Education and Research, Bhubaneswar 752050, Odisha, India}

\emailAdd{karthikin@tauex.tau.ac.il}
\emailAdd{sachin.jain@iiserpune.ac.in}
\emailAdd{vinaymm@iiserpune.ac.in}
\emailAdd{ abhishek.mehta@students.iiserpune.ac.in}
\emailAdd{pranjal.nayak@uky.edu}
\emailAdd{tarun\_sharma2@brown.edu, tarun.sharma@niser.ac.in}

\preprint{TAUP-3036\textbackslash19}

\abstract{We compute the two, three point function of the opearators in the spin zero multiplet of ${\cal N}=2$ Supersymmetric vector matter Chern-Simons theory at large $N$ and at all orders of 't Hooft coupling by solving the Schwinger-Dyson equation.  Schwinger-Dyson method to compute four point function becomes extremely complicated and hence we use bootstrap method to solve  for four point function of scaler operator $J_0^{f}=\bar\psi \psi$ and $J_0^{b}=\bar\phi \phi$.  Interestingly, due to the fact that $\langle J_0^{f}J_0^{f}J_0^{b} \rangle$ is a contact term, the four point function of  $ J_0^{f}$ operator  looks like that of free theory up to overall coupling constant dependent factors and up to some bulk AdS contact terms. On the other hand the $J_0^{b}$ four-point function receives an additional contribution compared to the free theory expression due to the $J_0^{f}$ exchange. Interestingly, double discontinuity of this single trace operator $J_0^{f}$ vanishes and hence it only contributes to AdS-contact term.}

%

\maketitle

\section{Introduction}
The perturbative technique to compute observables in quantum field theories involving Feynman diagrams 
is effective only when the coupling is weak and breaks down in the strong coupling regime. In the past few 
decades, various strong-weak dualities have been discovered proving to be extremely useful in understanding 
some of the most interesting non-perturbative properties of strongly coupled quantum field theories. One such 
class of dualities which have been studied extensively in recent times are the {\it Bosonization dualities} in 
Chern Simons gauge theories coupled to fundamental matter at large $N$ \cite{ Giombi:2011kc,Aharony:2011jz,Maldacena:2011jn,Maldacena:2012sf,Aharony:2012nh,Giombi:2016zwa}. Though, one of the main indications for these dualities initially 
came from their holographic duality with Vasiliev theories in $AdS_4$ \cite{Vasiliev:1990en,Vasiliev:1999ba,Klebanov:2002ja,Sezgin:2002rt,Giombi:2009wh,Giombi:2010vg}, by now there exists a plethora of evidence for these dualities coming from exact 
computations of correlation functions, thermal partition functions, scattering amplitudes and RG flow analysis 
relating these theories to known supersymmetric dualities \cite{Aharony:2012ns,Yokoyama:2012fa,Jain:2012qi,Yokoyama:2013pxa,Takimi:2013zca,Jain:2013py,Minwalla:2015sca,Gur-Ari:2015pca,Choudhury:2018iwf,Dey:2018ykx,Dey:2019ihe,Dandekar:2014era,Jain:2014nza,Inbasekar:2015tsa, Yokoyama:2016sbx,Inbasekar:2017ieo,Inbasekar:2017sqp,Maldacena:2011jn,Maldacena:2012sf,Aharony:2012nh,GurAri:2012is,Geracie:2015drf,Gur-Ari:2016xff,Yacoby:2018yvy}.

Primary example of these bosonization dualities are those among the quasi-fermionic
(critical bosonic and regular fermionic theory) and the quasi-bosonic theories (regular bosonic and critical 
fermionic theory)\footnote{In the terminology of \cite{Maldacena:2011jn,Maldacena:2012sf}}. 
A particularly interesting case of these dualities is present in the ${\cal N}=2$ supersymmetric $U(N)$ 
Chern Simons theory coupled to a single fundamental chiral multiplet. 
This theory exhibits a strong-weak self duality \cite{Benini:2011mf,Park:2013wta,Aharony:2013dha} generalizing
the well known Giveon-Kutosov duality \cite{Giveon:2008zn,Benini:2011mf}. 
The self duality of this supersymmetric theory serves as a parent duality for the non-supersymmetric 
bosonization dualities mentioned above since they can obtained from the supersymmetric theory via 
RG flows seeded by mass deformations \cite{Jain:2013gza,Aharony:2018pjn}. Taking hints from the supersymmetric 
dualities and the Level-Rank duality of pure Chern Simons theory, finite $N$ extensions for the non supersymmetric 
dualities have also been proposed \cite{Radicevic:2015yla,Aharony:2015mjs,Seiberg:2016gmd,Karch:2016sxi,Hsin:2016blu,Gomis:2017ixy,Cordova:2017vab,Cordova:2017kue,PhysRevB.95.205137,Cordova:2018qvg}. These theories were also investigated recently 
in presence of background magnetic field \cite{Halder:2019foo}. 

In this article, we will focus our attention on the ${\cal N}=2$ theory. Various large $N$ computations in this theory 
show remarkable features which are absent in the non supersymmetric couterparts. For example, the all loop $2\to 2$ 
scattering amplitude is tree-level exact except in anyonic channel \cite{Inbasekar:2015tsa} and it was shown that these 
amplitudes are also invariant under {\it Dual superconformal symmetry} \cite{Inbasekar:2015tsa,Inbasekar:2017sqp}. 
\cite{Inbasekar:2017ieo}  further showed that the tree level  $m \to n$ scattering amplitudes in this theory can be 
constructed using the BCFW recursions relations. 


Although many interesting non supersymmetric physical observables, as mentioned above, 
are amenable to direct exact computations by solving corresponding Dyson-Schwinger equations, the computation of 
4-point correlation function of even the simplest of single trace operators, namely the scalar operators 
$\bar\phi \phi$ and $\bar\psi \psi$, appears prohibitively difficult\footnote{See appendix \eqref{4ptcomp} for a 
discussion of our attempt.} to compute via this direct approach. Given the remarkable simplicity of the 
results for other known observables one expects the 4-point functions in this theory to also have a simple structure. 
In the present article, our main goal will be to determine the exact 2, 3 and leading connected 4-point correlation functions 
of scalar operators in this supersymmetric theory. 

For the quasi-bosonic and quasi-fermionic theories mentioned above, \cite{Turiaci:2018dht,Aharony:2018npf} 
recently determined the 4-point function of scalar operator using recently developed ideas from conformal Bootstrap. 
In particular, one of the central of the central objects used in \cite{Turiaci:2018dht} is the double discontinuity of the 
4-point function which determines the coefficients in the OPE expansion of external operators via the Lorenzian 
inversion formula (LIF) discovered by Caron-Huot in \cite{Caron-Huot:2017vep}. The authors of \cite{Turiaci:2018dht} 
first demonstrated that for large-$N$ CFTs the double discontinuity of the 4-point function of identical scalars 
determines the full 4-point function up to three $AdS_4$ contact Witten diagrams. The authors further showed 
that for the quasi-bosonic and quasi-fermionic theories the coefficients of these contact terms vanish. 
In the present work, we apply some of these ideas in conjunction with the self duality, to the case 
of scalar 4-point functions in our $N=2$ theory. 

Our article is structured as follows. In section \ref{review}, we describe the ${\cal N}=2$ theory of interest in this paper its 
operators spectrum in some detail. In section \ref{corrfn}, we determine the scalar multiplet 2 and 3-point functions via a 
directly computation. In section \ref{4ptfn}, we determine the 4-point function of the bosonic and the fermionic scalar 
operators in this theory using the double discontinuity technique developed in \cite{Turiaci:2018dht}. 
Finally, in the section \ref{discus}, we summarize our results and outline related open questions and future directions. 
In various appendix, we collect our notation and conventions, some technical details of the results in main text of the 
paper and briefly summarize our attempt at the direct computation of 4-point function. 

\paragraph{Note added in the proof :} While we were in the process of finishing up our article, we were informed about  
the related work \cite{Aharony:2019mbc} by the authors which has overlap with the results of our section \ref{corrfn}.

\section{${\cal N}=2$ theory and its Operator Spectrum}\label{review}
In this paper, we are interested in ${\cal N}=2$ $U(N)$ Chern-Simons theory coupled to single chiral multiplet, 
$\Phi \equiv (\phi,\psi)$, in the fundamental representation of the gauge group. The position space Lagrangian for 
the theory is
\begin{align}\label{susycs}
\mS_{\mN=2}^L = \int d^3x 
             \biggl[  & - \frac{\ka}{4\pi}\ep^{\mu\nu\rho}\text{Tr}\left( A_\mu\del_\nu A_\rho-\frac{2i}{3}A_\mu A_\nu A_\rho\right) 
            - \bps i \slashed{\mD}\psi  + \mD^\mu\bph\mD_\mu\phi\nn\\
&         + \frac{4\pi^2}{\ka^2}(\bph\phi)^3 - \frac{4\pi}{\ka}(\bph\phi)(\bps\psi) - \frac{2\pi}{\ka}(\bps\phi)(\bph\psi)\biggr]\ .
\end{align}

The theory above has two parameters : the rank of the gauge group, $N$, and the Chern-Simons level, $\kappa$, which 
is quantized to take only integer values \cite{Dijkgraaf:1989pz}. $\kappa^{-1}$ controls the strength of gauge interactions and the theory is 
perturbative for large values of $\kappa$ at any finite $N$.

This theory is conjectured to be self-dual under a strong-weak type duality, \cite{Benini:2011mf}. 
In the 't Hooft like large $N$ limit 
\be
\kappa \rightarrow \infty , N \rightarrow \infty \quad \textrm{with} \quad  \lambda = \frac{N}{\kappa} \quad \textrm{fixed} 
\ee
of interest in this paper, the duality transformation is
\begin{equation}\label{dualityst}
\kappa=-\kappa, \quad \lambda=\lam-\text{sgn}(\lambda)\ .
\end{equation}

Apart from the matching of many of the
supersymmetric observables which can be computed at finite $N$ and $\kappa$ using supersymmetric localization 
techniques, recent exact computation of many non-supersymmetric observables, e.g. the thermal partition function, 
in the  large $N$ limit \cite{Aharony:2011jz,Giombi:2011kc,Aharony:2012nh,GurAri:2012is,Jain:2013gza,Jain:2014nza} 
has provided ample evidence for this conjectured duality.


The theory is quantum mechanically (super) conformal for all values of $\kappa$ and $N$. In the 't Hooft limit, one 
can focus on the single trace superconformal primary operator spectrum of the theory. 
Though our theory has ${\cal N}=2$ superconformal symmetry, in this paper we will work in the ${\cal N}=1$ 
superspace formulation to allow us to use the relevant results of \cite{Inbasekar:2015tsa} for our 
computations. In the ${\cal N}=1$ language, the operators spectrum of the theory consists of a set of supercurrent 
operators \cite{Nizami:2013tpa}
\begin{equation}\label{eqJs}
J^{(s)}= \sum_{r=0}^{2s} (-1)^{\frac{r(r+1)}{2}} \begin{pmatrix} 2s \\ r \end{pmatrix} \nabla^r\bPh \nabla^{2s-r}\Phi~,
\end{equation}
which are written in terms of the superfields,$$\Phi = \phi+\theta \psi-\theta^2 F, \quad \bar\Phi = \bar\phi+\theta \bar\psi-\theta^2 \bar F~.$$
and the superscript $s$ in\eqref{eqJs} takes values in $\{ 0,\frac12,1,\frac32,\ldots \}$. Here, we have also defined
\be
\begin{split}
J^{(s)}=& \lam^{\al_1}\lam^{\al_2}\ldots \lam^{\al_{2s}} \mJ_{\al_1\al_2\ldots \al_{2s}}, \quad 
\nabla = \lam^{\al} \nabla_{\al} 
\end{split}
\ee
using the auxiliary commuting polarisation spinors, $\lam^{\al_i}$, which keep track of the spin; and $\nabla_{\al_i}$ are 
the standard supersymmetry invariant gauge-covariant derivatives. Their action on the matter superfields of our theory is given by,
\begin{align}
    \nabla_\al\Phi=& D_\al\Phi-i\Ga_\al \Phi \nn \\
    \nabla_\al\bPh=& D_\al\bPh+i \Ga_\al \bPh
\end{align}
The explicit expressions for the first few spin-$s$ currents are,
\begin{equation}\begin{aligned}
     J_0 & = \bPh\Phi\\
    J_\al &= \bPh \na_\al \Phi - \na_\al\bar{\Phi} \Phi=\bPh D_\al \Phi - D_\al \bPh \Phi - 2i \bPh \Ga_\al \Phi\\
    J_{\al\beta} &=\bPh\na_\al\na_\beta\Phi-2 \na_\al\bPh\na_\beta\Phi+\na_\al\na_\beta\bPh\Phi\\
    J_{\al\beta\ga} & =\bPh \na_\al\na_\beta\na_\ga\Phi-3\na_\al\bPh\na_\beta\na_\ga\Phi-3\na_\al\na_\beta\bPh \na_\ga\Phi+\na_\al\na_\beta\na_\ga\bPh \Phi
\end{aligned}\end{equation}
In the free limit of the theory i.e. $\lambda \rightarrow 0$, each of these supercurrents, ${\cal J}^{(s)}$ with $s\neq 0$, 
satisfies the conservation equation

\begin{equation}\label{conseqn}
{\cal D}^\A \left( \frac{\partial}{\partial \lambda^\A} {\cal J}^{(s)} \right) = 0 
\end{equation}
 and constitutes two  component conserved current operators $\{ J^{(s)}, J^{(s+\half)} \}$ in its $\theta$ expansion \cite{Nizami:2013tpa}. 
 At finite $\lambda$, the conservation equation \eqref{conseqn} is violated at order $\frac{1}{N}$ by double trace operators 
 \cite{Nizami:2013tpa,Giombi:2011kc}. 

In this article, we are interested in the scalar operator ${\cal J}^{(0)}(\theta,x)$. There is no conservation 
equation associated with this operator and it constitutes 2 scalar and 1 spin half operator as follows
\begin{align} 
J^{(0)}(\theta,x) = J_0^b(x) + \theta^\al \Psi_{\al}(x) - \theta^2 J_0^f(x) \
\end{align}
where 
\begin{equation}
J_0^b(x) =\bph \phi(x)  , \quad   \Psi_{\al}(x)= (\bph \psi_\al + \bps_\al\phi)(x) , \quad  J_0^f(x) = \bps \psi(x).
\end{equation}
In the subsequent sections, we compute the 2 and 3-point functions of the $J^{(0)}$ operator and two 
component of the 4-point function.

\section{Correlation functions}\label{corrfn}
\begin{figure}
\begin{center}
\includegraphics[width=10cm,height=5cm]{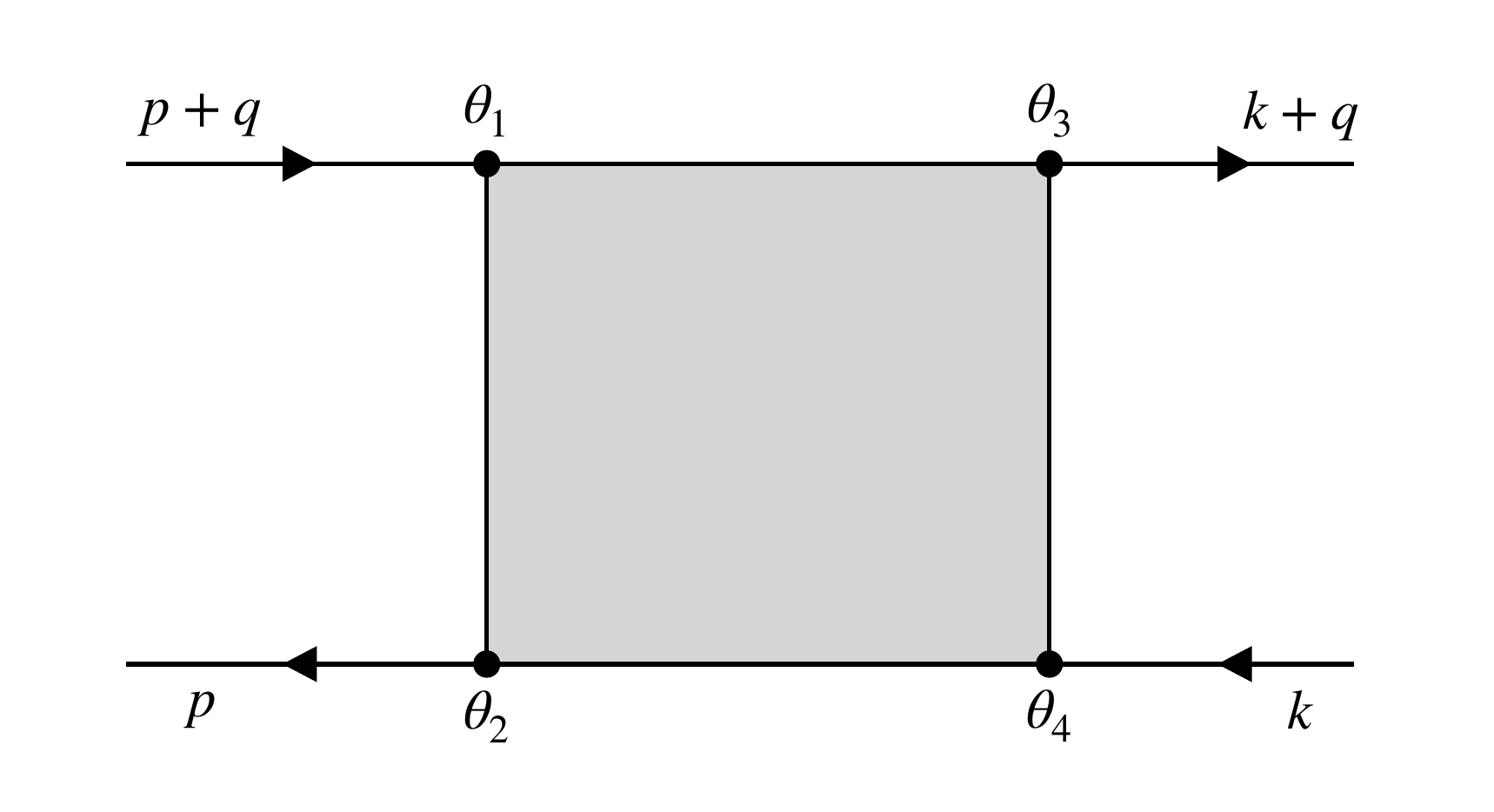}
\caption{Diagrammatic representation of the exact four point vertex, $\nu_4$ in \eqref{nu4vert}. \label{nu4diag}}
\end{center}
\end{figure}

In this section, we compute the two and three point correlation function of the
$J_{0}(\theta,p)$ operator. Two of the main ingredients for these computations are the exact propagator (\ref{propag})
and the renormalized four point vertex for the fundamental superfield $\Phi(\theta,p)$ ($\nu_4$ in \ref{nu4vert}). These were computed in
\cite{Inbasekar:2015tsa} for a more general class of theories with ${\cal N}=1$ supersymmetry which can be 
thought of as one parameter\footnote{Quartic superpotential term : 
$-\frac{\pi(\omega-1)}{\kappa} \int d^3x ~d^2\theta ~(\bar\Phi \Phi)^2$. $\omega=1$ is the ${\cal N}=2$ point.} 
deformation of the ${\cal N}=2$ theory of interest in this paper. 
Below, we list these results for our ${\cal N}=2$ theory, conveniently stated in term of the exact quantum effective 
action 
\be
\begin{split}\label{QEA}
S &= S_2 + S_4 ~, \\
S_2 &= \int \frac{d^3p}{(2\pi)^3} d^2\theta_1 d^2\theta_2  \left[ \bar{\Phi}(\theta_1,-p)  e^{-\theta_1^\A p_{\A\B} \theta_2^\B}  
           \Phi(\theta_2,p) \right] , \\
S_4 &= \frac{1}{2} \int \frac{d^2p}{(2\pi)^3} \frac{d^2q}{(2\pi)^3} \frac{d^2k}{(2\pi)^3} 
                            d^2\theta_1 d^2\theta_2 d^2\theta_3 d^2\theta_4  \\ 
& \hspace{1.5cm}  \bigg[ \nu_4(\theta_1,\theta_2,\theta_3,\theta_4;p,q,k)  
                             \Phi_i(\theta_1,-(p+q)) \bar{\Phi}^i(\theta_2,p) \bar{\Phi}^j(\theta_3,k+q) \Phi_j(\theta_4,-k) \bigg] \\ 
\end{split}
\ee
The quadratic part of the effective action receives no quantum corrections at large $N$ in the ${\cal N}=2$ theory. 
The propagator is thus tree level exact and given by 
\be
\begin{split}\label{propag}
\la \bar\Phi(\theta_1,p_1) \Phi(\theta_2,p_2) \ra &= (2\pi)^3 \delta^3(p_1+p_2) {\cal P}(\theta_1,\theta_2; p_1) \\
   & = (2\pi)^3 \delta^3(p_1+p_2) \frac{e^{-\theta^\A_1 \theta^\B_2 (p_1)_{\A\B}}}{p_1^2}.
\end{split}
\ee
The quartic superspace vertex, $\nu_4$, does receive quantum corrections and takes the following form
\be\begin{aligned}\label{nu4vert}
\nu_4(\theta_1,\theta_2,\theta_3,\theta_4;p,q,k) & = e^{\frac{1}{4} X.(p.X_{12}+q.X_{13}+k.X_{43})} F_4(X_{12}, X_{13}, X_{43}; p,q,k), \\
\textrm{with} \quad  F_4 &= X_{12}^+ X_{43}^+ \bigg[ A(p,q,k) X_{12}^- X_{43}^- X_{13}^+ X_{13}^-  + C(p,q,k) X_{12}^- X_{13}^+   \\
               &\hspace{5cm}+ D(p,q,k) X_{13}^+ X_{43}^-   \bigg]
\end{aligned}\ee
The overall exponential factor is determined by supersymmetric Ward identity \eqref{WIder}, while the coefficient functions 
A, C and D require explicit computation and given by \cite{Inbasekar:2015tsa}
\be
\begin{split}\label{ACD}
A(p,q,k) & = -\frac{2\pi i}{\kappa} e^{2i\lambda \left[ \tan^{-1} \left( \frac{2k_s}{q_3} \right) - \tan^{-1} \left( \frac{2p_s}{q_3} \right) \right] },  \\
C(p,q,k) & =   D(p,q,k) = \frac{2 A(p,q,k)}{(k-p)_-}. \\
\end{split} 
\ee
Note that the vertex $\nu_4$ was computed in a special momentum configuration, namely 
\be\label{collimit}
q_+ = q_- = 0.
\ee
while the momenta $p$ and $k$ are arbitrary \footnote{We refer the reader to appendix \ref{NandC} for conventions 
for labelling momenta.}. For this reason, our computation of correlation functions will 
also be restricted configuration in which the momentum of $J_0$ operators are restricted to lie only in the 3-direction. 
Diagrammatically, the exact four point vertex will be represented as in Figure \ref{nu4diag}.

\subsection{Constraints on correlation functions from supersymmetry}\label{WardIdn}
To begin with, let us study the constraints on an arbitrary correlation function due to supersymmetry. As stated earlier, although our theory has ${\cal N}=2$ supersymmetry, we will be working in ${\cal N}=1$ superspace following \cite{Inbasekar:2015tsa}. A general n-point correlation function of ${\cal N}=1$ scalar superfield is 
constrained by supersymmetry and translation invariance to take the following form \cite{Inbasekar:2015tsa}
\be
\begin{split}\label{widen}
& \la {\cal O}_1(\th_1,p_1) \ldots {\cal O}_n(\th_n,p_n) \ra \\
   &  \qquad = (2\pi)^3 \delta^3 \left( \sum_{i=1}^n p_i \right) \exp \left[ \left( \frac{1}{n}\sum_{i=1}^n \th_i \right).\left( \sum_{i=1}^n p_i.\th_i \right) \right]
                   F_n( \{\th_{in}\} ; \{ p_i \})
\end{split}
\ee
where $\th_{ij}=\th_i - \th_j$. 
The $\delta^3(\sum_i p_i)$ follows from translation invariance while the overall Grassmann exponential factor 
follows from invariance under ${\cal N}=1$ supersymmetry. Note that the function $F_n$ above only depend on the 
differences of the Grassmann coordinates. Following \cite{Inbasekar:2015tsa}, the form is 
easily derived as follows
\be
\begin{split}\label{WIder}
0 &= \left[ \sum_{i=1}^{n} {\cal Q}^{(i)}_\A \right] \la {\cal O}_1(\th_1,p_1) \ldots {\cal O}_n(\th_n,p_n) \ra \\
  & = \left[ \sum_{i=1}^{n} \left( \frac{\del}{\del \theta_i^\A} - (p_i)_{\A\B} \theta_i^\B \right) \right] \la {\cal O}_1(\th_1,p_1) \ldots {\cal O}_n(\th_n,p_n) \ra \\
  & = \left( n \frac{\del}{\del X^\A} - \sum_{i=1}^{n-1} (p_i)_{\A\B} X_{in}^\B \right) 
                     \la {\cal O}_1(\th_1,p_1) \ldots {\cal O}_n(\th_n,p_n) \ra \\
\end{split}
\ee
In the last line above, we used the momentum conservation to replace $p_n$ with $ \sum_{i=1}^{n-1} (-p_i)$. Defining the sum 
and the difference of  Grassmann variables as
\be
X^\A = \sum_{i=1}^n \theta_{i}^\A \quad , \quad X_{in}^\A = \theta_i^\A - \theta_n^\A.
\ee
The factorised form in \eqref{widen} follows as the solution to last equation in \eqref{WIder}.


\subsection{\texorpdfstring{$J_0$}{J0}-vertex}\label{J0vertex}
Before proceeding to the computation of correlation functions, it would be useful to compute an intermediate quantity, 
the $J_0$-vertex.  It is defined by stripping of the propagators from $\la J_0 \Phi \bar\Phi \ra$ as follows 
\be
\begin{split}
&  \la J_0(\theta_1,p_1) \Phi(\theta_2,p_2) \bar\Phi(\theta_3,p_3) \ra = \\
&  \quad   \int \frac{d^3p'_2}{(2\pi)^3} \frac{d^3p'_3}{(2\pi)^3} d^2\theta'_2  d^2\theta'_3 
               ~\left[ \la J_0(\theta_1,p_1) \Phi(\theta'_2,p'_2) \bar\Phi(\theta'_3,p'_3) \ra_{ver} 
                 {\cal P}(\theta'_2,\theta_2; -p_2) {\cal P}(\theta'_3,\theta_3; p_3) \right] \\
\end{split}
\ee
and satisfies the same Ward identity as a three point function \eqref{widen}. 

\begin{figure}
\begin{center}
\includegraphics[width=15cm,height=5cm]{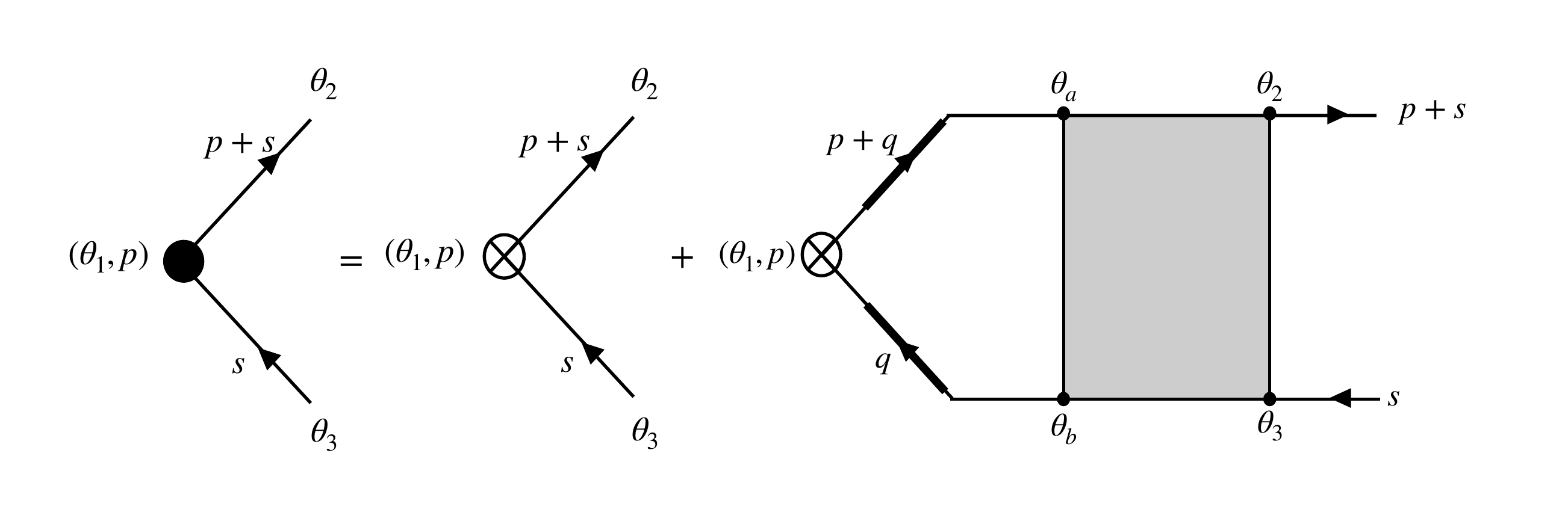}
\caption{Solid circle on the LHS represents the full exact $J_0$ vertex and the first diagram on RHS is the free vertex. 
The second diagram on RHS includes all the interactions which are accounted by insertion of exact 4 point vertex \eqref{nu4vert} 
connected to the external ${\cal J}^{(0)}$ operator using the exact propagator. \label{nu3diag} }
\end{center}
\end{figure}

The vertex receives contribution 
both from the free propagation of the fundamental field as well as from the interaction vertices in the theory. The free part 
vertex is simply proportional to the momentum and the Grassmannian  $\delta$-functions while the interacting part of the 
vertex can be computed from the exact $\nu_4$ vertex. Figure \eqref{nu3diag} shows the relevant diagrams.
\be
\begin{split}\label{J0vert}
& \la J_0(\theta_1,p) \Phi(\theta_2,r) \bar{\Phi}(\theta_3,s) \ra_{ver}  =  \la J_0(\theta_1,p) \Phi(\theta_2,r) \bar{\Phi}(\theta_3,s) \ra_{ver,free}  
                            + \la J_0(\theta_1,p) \Phi(\theta_2,r) \bar{\Phi}(\theta_3,s) \ra_{ver,int}  \\
& \\
& \textrm{where} \quad \la J_0(\theta_1,p) \Phi(\theta_2,r) \bar{\Phi}(\theta_3,s) \ra_{ver,free} 
                   = (2\pi)^3\delta^3(p+r+s) \ \ \nu_{3,free}(\theta_{12}, \theta_{32}; p,s)  \\
&  \hspace{6.3cm}   = (2\pi)^3\delta^3(p+r+s) \theta_{32}^+ \theta_{32}^- \theta_{12}^+ \theta_{12}^-  \\
& \textrm{and} \quad \la J_0(\theta_1,p) \Phi(\theta_2,r) \bar{\Phi}(\theta_3,s) \ra_{ver,int}  \\
& \qquad           = (2\pi)^3\delta^3(p+r+s) \left[ \int \frac{d^3q}{(2\pi^3)} d^2\theta_a d^2\theta_b ~
                            {\cal P}(\theta_1,\theta_a; q+p) {\cal P}(\theta_b,\theta_1; q) 
                            \nu_4(\theta_a,\theta_b,\theta_2,\theta_3 ; q,p,s) \right] \\
& \qquad          = (2\pi)^3 \delta^3(p+r+s) e^{\frac{1}{3}\theta_{123}.(p.\theta_{12}+s.\theta_{32})} 
                    \nu_{3,int}(\theta_{12},\theta_{32},p,s) \\
\end{split}
\ee
Explicit computation of the above integral, with constraint $p_+ = p_- = 0$ following from the \eqref{collimit}, 
leads to the following result for the full $J_0$-vertex factor
\be
\begin{split}
\nu_3 &= \left( \nu_{3,free}+\nu_{3,int} \right) (\theta_{12},\theta_{32},p,s) \\
&  = \frac{1}{2s^+} \left[ 1-e^{2i \lambda \tan^{-1}(\frac{2 s_s}{p_3})} \right] \theta_{32}^+ \theta_{12}^+ 
     +  \frac{1}{2p_3} \left( e^{2i \lambda \tan^{-1}(\frac{2 s_s}{p_3})  - i\pi\lambda sgn(p_3) } - 1 \right) \theta_{32}^+ \theta_{32}^- \\
&  \qquad + \left( 1 + \frac{1}{6} (-4 + e^{2i \lambda \tan^{-1}(\frac{2 s_s}{p_3})} + 3 e^{2i \lambda \tan^{-1}(\frac{2 s_s}{p_3}) 
                 - i\pi\lambda sgn(p_3)}) \right)  \theta_{32}^+ \theta_{32}^- \theta_{12}^+ \theta_{12}^- 
\end{split}
\ee

The $J_0$-vertex computed above will be useful in further computations of 2 and 3 point functions of the $J_0$ 
operator.

\subsection{\texorpdfstring{$\langle J_0 J_0 \rangle$}{<J0J0>} correlation function} \label{J02pt}
\begin{figure}
\begin{center}
\includegraphics[width=10cm,height=5cm]{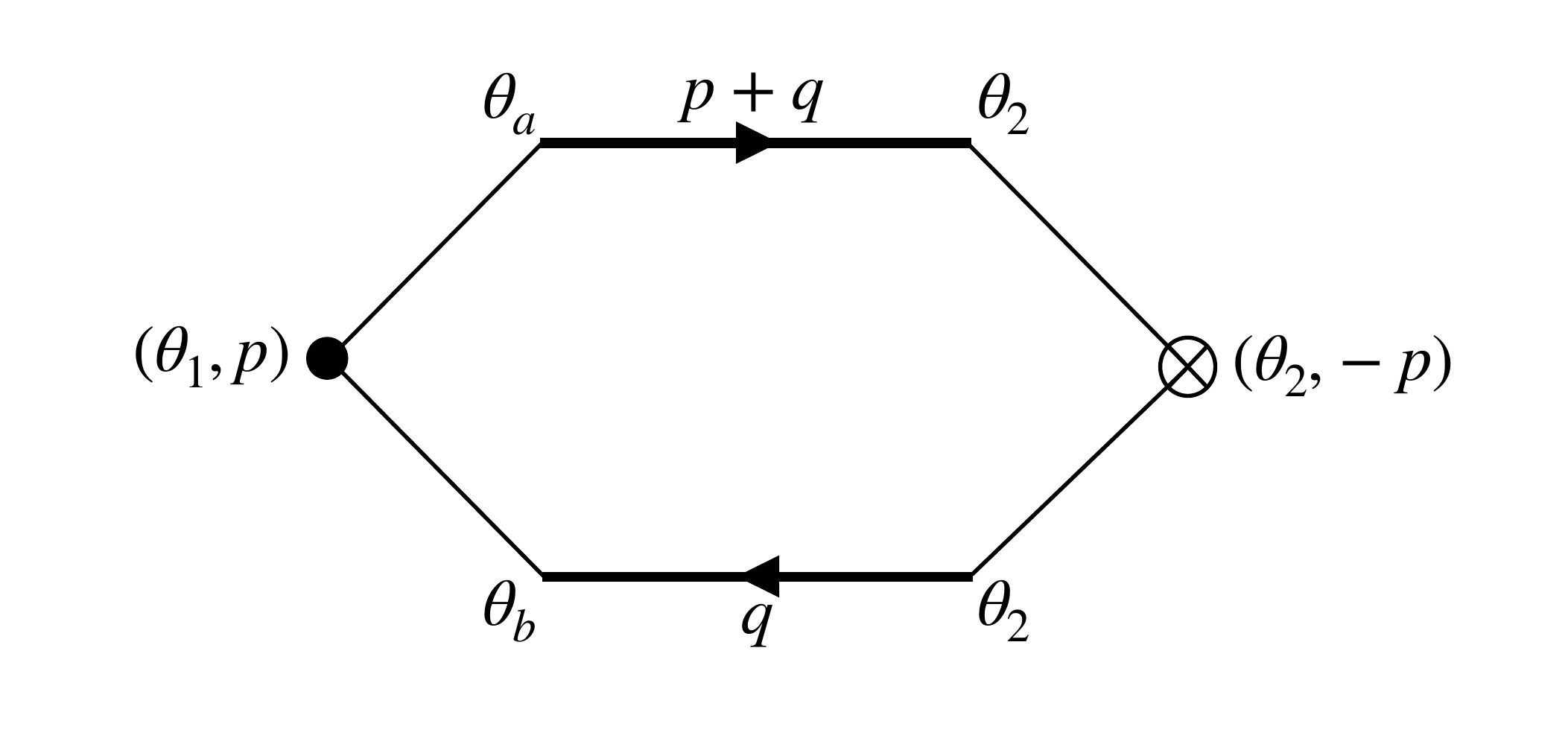}
\caption{The full $J_0$ 2 point function is obtained by connecting the exact $J_0$-vertex (solid circle) to the free vertex 
(cross) with exact propagators (thick line).\label{J02ptdiag}}
\end{center}
\end{figure}

The 2 point function can be straightforwardly computed from the $J_0$-vertex determined in the previous section 
by combining the exact vertex on one side with the free vertex on the other side. Figure \eqref{J02ptdiag} shows the 
relevant diagram which leads to the following integral for the two point function
\be\label{J02ptfn}
\begin{split}
 & \la J_0(\theta_1,p)J_0(\theta_2,r) \ra \\
 & \quad = (2\pi)^3 \delta^3(p+r) \left[ N \int \frac{d^3q}{(2\pi)^3} d^2\theta_a d^2\theta_b 
                 ~\nu_3(\theta_1,\theta_a,\theta_b ; p,-p-q,q) {\cal P}(\theta_a,\theta_2; q+p) {\cal P}(\theta_2,\theta_b; q) \right] \\
\end{split}
\ee
Again, the collinear constraint \eqref{collimit} restricts the momenta $p$ and $s$ to lie in 3-direction. 
Computing the integrals with this constraint leads to the following result
\be
\begin{split}
& \la J_0(\theta_1,p)J_0(\theta_2,r) \ra \\
& \qquad  = (2\pi)^3 \delta^3(p+r) N \frac{e^{-\theta_1.p.\theta_2}}{8 |p_3|} 
    \left( \frac{\sin(\pi\lambda)}{\pi\lambda} + |p_3| \delta^2(\theta_{12}) \frac{1-\cos(\pi\lambda)}{\pi\lambda} \right) \\
\end{split}
\ee
The result can be straightforwardly generalized for arbitrary external momenta to give \\
\be
\begin{split}\label{J02ptresult}
& \la J_0(\theta_1,p)J_0(\theta_2,r) \ra \\
& \qquad = (2\pi)^3 \delta^3(p+r) N \frac{e^{-\theta_1.p.\theta_2}}{8 |p|} 
    \left( \frac{\sin(\pi\lambda)}{\pi\lambda} + |p| \delta^2(\theta_{12}) \frac{1-\cos(\pi\lambda)}{\pi\lambda} \right) \\
\end{split}
\ee

The non vanishing component correlators can easily be read off to give
\be
\begin{split}\label{J02ptcomp}
& \langle J^b_0(p) J^b_0(-p)\rangle = \frac{N}{8 |p|} \frac{\sin(\pi\lm)}{\pi\lm} \\
& \langle J^f_0(p) J^f_0(-p)\rangle = - \frac{N|p|}{8} \frac{\sin(\pi\lm)}{\pi\lm} \\ 
& \langle\Psi_\A(p) \Psi_\B(-p)\rangle = \frac{N}{8} \left( \frac{p_{\A\B}}{|p|} \frac{\sin(\pi\lm)}{\pi\lm} 
                        + C_{\A\B} \frac{1-\cos(\pi\lm)}{\pi\lm} \right) \\ 
& \langle J^b_0(p) J^f_0(-p)\rangle = - \frac{N}{8} \frac{(1-\cos(\pi\lm))}{\pi\lm} \\
\end{split}
\ee

Let us compare the above two-point functions with the corresponding two-point functions in 
the regular fermionic and regular bosonic theories studied in \cite{Aharony:2012nh} and \cite{GurAri:2012is} 
respectively. 

Note that as opposed to the regular bosonic and regular fermionic theories studied 
in \cite{Aharony:2012nh} and \cite{GurAri:2012is}, the $\lambda$ dependence of the two-point function 
of $J^b_0$ and $J^f_0$ operators is the same as that of the higher spin currents 
in the non-supersymmetric cases. Further, using the double trace factorization argument of 
\cite{Gur-Ari:2015pca} relating the two-point function of current operators 
in the supersymmetric and the above mentioned non-supersymmetric theories, we know 
that the two-point function of all the current operators in our supersymmetric theory is 
exactly the same as those of the corresponding regular boson/fermion theory. Thus, we see that 
in our theory the two-point function of scalar operators is the same as that for 
the higher spin current operators. The reason for this is supersymmetry. 
Though we are working in ${\cal N}=1$ superspace language, our theory has underlying 
${\cal N}=2$ supersymmetry under which the scalar operators $J^b_0, J^f_0$ belong to the  
same supersymmetry multiplet as the spin 1 conserved current and thus the two-point function of 
the two are thus related by supersymmetry. 


\subsection{\texorpdfstring{$\langle J_0 J_0 J_0 \rangle$}{<J0 J0 J0>} correlation function}\label{J03pt}
\begin{figure}
\begin{center}
\includegraphics[width=15cm,height=5cm]{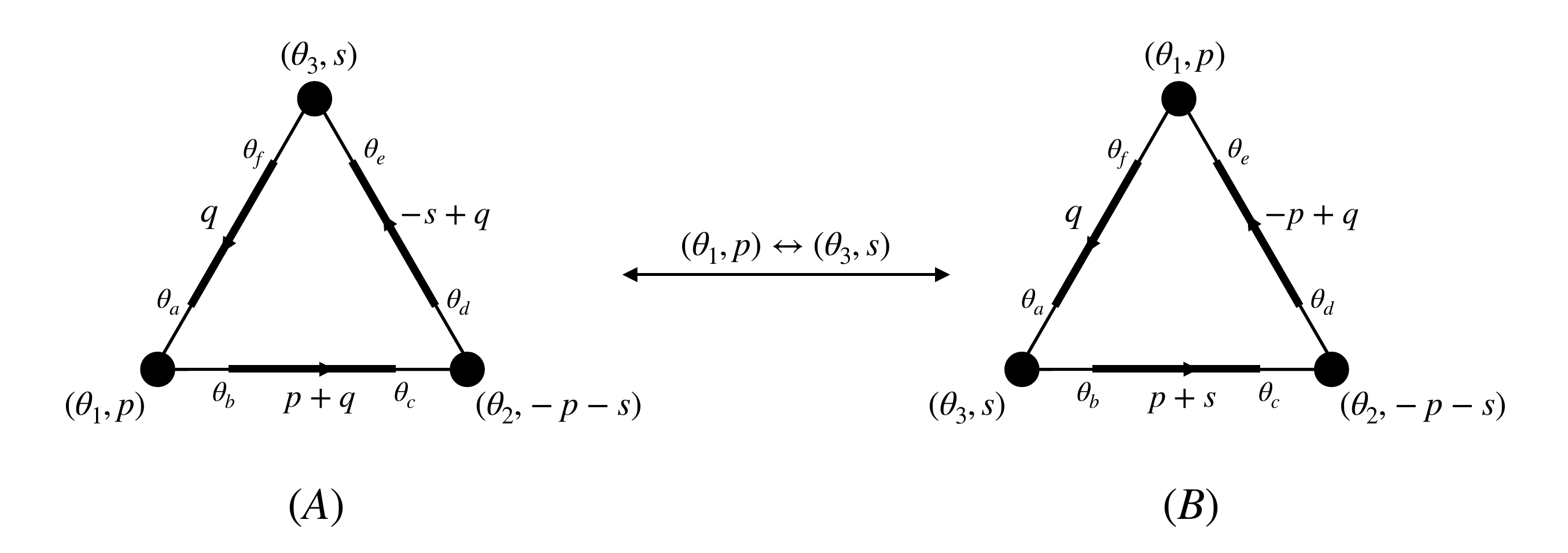}
\caption{The full $J_0$ 3 point function is obtained by connecting three exact $J_0$-vertices with exact propagators. 
There are two such diagrams, as shown above, which turn out to be equal.\label{J03ptdiag}}
\end{center}
\end{figure}

The full 3-point function can be constructed by combining three $J_0$ vertices with exact propagators. 
There are two such diagrams shown in figure \eqref{J03ptdiag}. 
Each of these two diagrams can easily be shown to be cyclically symmetric and related to each other 
by pair-exchange of any two $J_0$ insertions. An explicit computation of the diagram shows that each of the diagrams 
is completely symmetric (cyclic as well as under pair-exchange) by itself and the two diagrams are equal. 
The full 3 point function is then just twice the contribution of the first diagram which we write down below.
\be
\begin{split}\label{J03pteqn}
& \langle J_0(\theta_1,p) J_0(\theta_2,r) J_0(\theta_3,s)\rangle  = (2\pi)^3 \delta^3(p+r+s) G_3(\theta_1,\theta_2,\theta_3 ; p,s) , \\
& \textrm{where} \\
& \quad G_3(\theta_1,\theta_2,\theta_3 ; p,s) = 
            2 N \int \frac{d^3q}{(2\pi)^3} \left( \prod_{i=a}^{f} d^2\theta_i \right) 
            \bigg[ \nu_3(\theta_1,\theta_a,\theta_b ; p,-p-q) \\
            &  \hspace{4cm} \times \nu_3(\theta_2,\theta_c,\theta_d ; -p-s,s-q) \ \nu_3(\theta_3,\theta_e,\theta_f ; s,-q) \\
            & \hspace{4cm} \times {\cal P}(\theta_a, \theta_f ; -q) {\cal P}(\theta_a, \theta_f ; s-q) {\cal P}(\theta_c, \theta_b ; -p-q) \bigg]  \\
\end{split}
\ee
The overall factor of 2 in the above equation is from the sum over two triangle diagrams in figure \eqref{J03ptdiag} which turn 
out to be equal while the factor of $N$ results from index contractions. Explicit computation of the above integrals in the collinear 
limit of the external momenta gives the following result
\be
\begin{split}\label{J03ptresult}
& G_3(\theta_1,\theta_2,\theta_3 ; p,s)  =  e^{\frac{1}{3}\theta_{123}.(p.\theta_{12}+s.\theta_{32})} 
                                     F_3(\theta_{12},\theta_{32},p,s) , \\
& \quad  F_3(\theta_{12},\theta_{32},p,s) = 2 \left( A_1 + A_2 \theta_{12}^+ \theta_{12}^- + A_3 \theta_{32}^+ \theta_{32}^- + 
                   A_4 \theta_{12}^+ \theta_{32}^- + A_5 \theta_{32}^+ \theta_{12}^- + A_6 \theta_{12}^+ \theta_{12}^- \theta_{32}^+ \theta_{32}^-  \right) \\
\end{split}
\ee
The overall factor of 2 in the expression of $F_3$ above is from the sum over two triangle diagrams 
which turn out to be equal. The coefficients $\{ A_i \}$ are given by 
\be
\begin{split}\label{F3coeffs}
&  A_1 =  \frac{\sin(2\pi\lambda)}{2\pi\lambda} \frac{1}{8 |p_3| |s_3| |p_3 + s_3|} , \\
&  A_2 = - i  \frac{(\sin(\pi\lambda))^2}{\pi\lambda} \frac{1}{8 |s_3| |p_3 + s_3|} , \\
&  A_3 = - i \frac{(\sin(\pi\lambda))^2}{\pi\lambda} \frac{1}{8 |p_3| |p_3 + s_3|} , \\
&  A_4 = - \frac{1}{48 p_3 s_3 (p_3+s_3)} 
    \bigg[ \frac{\sin (2 \pi  \lambda)}{2 \pi  \lambda} \bigg( -(p_3+2 s_3) \text{sgn}(p_3)+(2 p_3+s_3)
   \text{sgn}(s_3)+(p_3-s_3) \text{sgn}(p_3+s_3) \bigg) \\
& \hspace{3.5cm}   - 3 i  \frac{\sin ^2(\pi  \lambda)}{\pi  \lambda }  \text{sgn}(p_3+s_3) \bigg( |p_3+s_3|-(|p_3| + |s_3|) \bigg) \bigg] \\
&  A_5 = \frac{1}{48 p_3 s_3 (p_3+s_3)} 
    \bigg[ \frac{\sin (2 \pi  \lambda)}{2 \pi  \lambda} \bigg( -(p_3+2 s_3) \text{sgn}(p_3)+(2 p_3+s_3)
   \text{sgn}(s_3)+(p_3-s_3) \text{sgn}(p_3+s_3) \bigg) \\
& \hspace{3.5cm}    + 3 i  \frac{\sin ^2(\pi  \lambda)}{\pi  \lambda }  \text{sgn}(p_3+s_3)  \bigg( |p_3+s_3|-(|p_3| + |s_3|) \bigg) \bigg] \\
&  A_6 = - \frac{\sin(2\pi\lambda)}{2\pi\lambda} \bigg[ \frac{1}{72 p_3 s_3 (p_3+s_3)} 
      \bigg( (p3-s3)(2 p3+s3) \text{sgn}(p_3) - (p_3 + 2 s_3)  (p_3 - s_3) \text{sgn}(s_3) \\ 
& \hspace{6cm}  - (p_3 + 2 s_3) (2 p_3 + s_3) \text{sgn}(p_3 + s_3)  \bigg) \bigg] 
\end{split}
\ee 
The non vanishing components of the three point functions can easily be extracted from \eqref{J03ptresult} and \eqref{F3coeffs} to be
\be
\begin{split}\label{3ptcompcorr}
& \langle J^b_0(p_3) J^b_0(s_3) J^b_0(-p_3-s_3)\rangle =  \frac{\sin(2\pi\lambda)}{2\pi\lambda} \frac{1}{8 |p_3 s_3 (p_3+s_3)|} \\
& \langle J^f_0(p_3) J^f_0(s_3) J^f_0(-p_3-s_3)\rangle = - \frac{i}{8} \frac{(\sin(\pi\lambda))^2}{\pi\lambda}  \\
& \langle J^b_0(p_3) J^b_0(s_3) J^f_0(-p_3-s_3)\rangle = \frac{(\sin(\pi\lambda))^2}{\pi\lambda} \frac{-i}{8 |p_3 s_3|} \\
& \langle J^f_0(p_3) J^f_0(s_3) J^b_0(-p_3-s_3)\rangle =  \frac{\sin(2\pi\lambda)}{2\pi\lambda} \frac{1}{16 |p_3+s_3|} \\
& \langle\Psi_+(p_3) \Psi_-(s_3) J^b_0(-p_3-s_3)\rangle =  \frac{1}{16 p_s s_3(p_3+s_3)} 
              \bigg( \frac{\sin(2\pi\lambda)}{2\pi\lambda} \big( |p3|-|s3|-(p3-s3) \text{sgn}(p3+s3) \big) \\ 
&   \hspace{7.5cm}   - i \frac{(\sin(\pi\lambda))^2}{\pi\lambda} \text{sgn}(p3+s3) (|p3+s3|-|p3|+|s3|) \bigg) \\ 
& \langle\Psi_+(p_3) \Psi_-(s_3) J^f_0(-p_3-s_3)\rangle =  \frac{1}{16 p_s s_3} 
              \bigg( \frac{\sin(2\pi\lambda)}{2\pi\lambda} \big( |p3 + s3| - |p3| - |s3| \big) \\ 
&   \hspace{5cm}   + i \frac{(\sin(\pi\lambda))^2}{\pi\lambda} \text{sgn}(p3+s3) \big( (p3 - s3) |p3 + s3| - |p3| + |s3| \big)  \bigg) \\ 
\end{split}
\ee

Notice that in the above result for 3 point functions, two different functional forms of $\lambda$ dependences appear, 
namely $\frac{\sin \pi\lambda}{\pi\lambda}$ and $\frac{\sin^2\pi\lambda}{\pi\lambda}$. The two of them differ in a crucial 
way. The first one has a finite $\lambda \rightarrow 0$ limit and is invariant under parity under which $\lambda$ is odd. 
The second is odd under parity and vanishes in $\lambda \rightarrow 0$ limit. This result thus provides some 
support for the conjecture made in \cite{Nizami:2013tpa} that the three-point functions in ${\cal N}=1$ superconformal 
theories with higher spin symmetry have exactly one parity even and one parity odd structure. 
The results \eqref{J02ptresult} and \eqref{F3coeffs} for the 2 and 3-point are clearly invariant under the duality 
transformation \eqref{dualityst}.

\section{Four point functions}\label{4ptfn}
In the previous section, we evaluated 
the 3-point functions involving the ${\cal J}_0$ operator in the $\mathcal{N}=2$ supersymmetric theory by computing 
the required vertex. However, the direct computation of the four-point function of  $J_0$  operator following the same 
technique has proven to be intractable in our attempt till now. We describe our attempt to evaluate this four-point 
function in momentum space through the required vertices in the Appendix (\ref{4ptcomp}). 

In this section, we determine the four-point correlators of the $J_0^b$ and $J_0^{f}$ operators using a novel 
method developed in \cite{Turiaci:2018dht}, which we briefly review below. Note that we will be evaluating the 
4-point correlation function in the position space as in \cite{Turiaci:2018dht}.

Consider the position space four-point correlator of the identical external operators with conformal dimensions 
$\Delta$. The function $\mathcal{A}$ which is known as the reduced correlator is defined as follows
\begin{align} 
\langle\mathcal{O}(x_1)\mathcal{O}(x_2)\mathcal{O}(x_3)\mathcal{O}(x_4)\rangle = \frac{1}{x^{2\Delta}_{12}}\frac{1}{x^{2\Delta}_{34}}\mathcal{A}(u, v) = \frac{1}{x^{2\Delta}_{13}}\frac{1}{x^{2\Delta}_{24}}\frac{\mathcal{A}(u, v)}{u^{\Delta}}~.
\end{align}
Here, $u,v$ are the standard cross-ratios:$$u = \left(\frac{|x_{12}||x_{34}|}{|x_{13}||x_{24}|}\right)^2, \quad v=\left(\frac{|x_{14}||x_{23}|}{|x_{13}||x_{24}|}\right)^2~.$$The conformal block expansion expressed in terms of the reduced correlator $\mathcal{A}(u, v)$ is given as
\begin{align}\label{RC}
\frac{\mathcal{A}(u, v)}{u^{\Delta}} = \frac{1}{u^{\Delta}}\sum_{k} C^2_{\mathcal{O}\mathcal{O}O_k}G_{\Delta_k, J_k}(u, v)
\end{align}
 where $G_{\Delta_k, J_k}(u, v)$ is known as the conformal block corresponding to the operator $\mathcal{O}_k$ with scaling dimension $\Delta_k$ and spin $J_k$. 

\begin{figure}[h!]
\begin{center}
 \includegraphics[width=6cm,height=4cm]{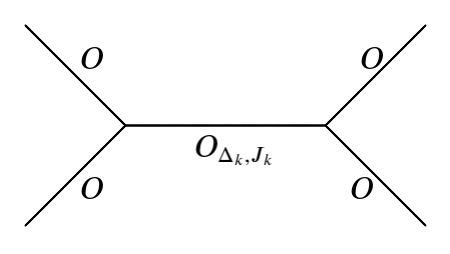}
  \caption{Schematic for the conformal block expansion}
  \end{center}
\end{figure}
In the supersymmetric four point functions of $J_0$ operators, the relevant exchanges are schematically shown below

\begin{figure}[h!]
\begin{center}
 \includegraphics[width=6cm,height=4cm]{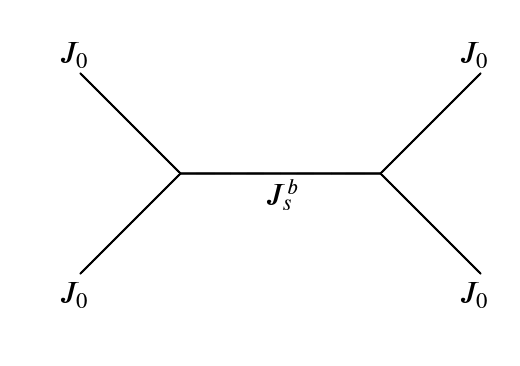}
  \includegraphics[width=6cm,height=4cm]{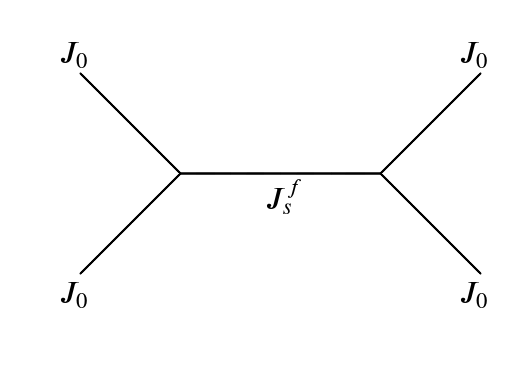}
  \caption{Schematic for the exchanges relevant in the supersymmetric scalar correlators }
  \end{center}
\end{figure}

\subsection{Review of the double discontinuity technique}
In \cite{Turiaci:2018dht}, the authors determine the four-point correlation functions of the scalar operator in the non-supersymmetric scalar/fermion coupled to Chern Simons gauge field i.e. quasi-bosonic and quasi-fermionic theory respectively. In order to obtain the required four-point functions, the authors utilize the inversion formula which relates the double discontinuity to the OPE coefficients \cite{Caron-Huot:2017vep}.  The authors first prove an interesting theorem that in the large-N limit of a $CFT_d$, the double discontinuity constrains the four-point correlator up to three contact terms in $AdS_{d+1}$. Suppose there are two solutions  $G_1$ and $G_2$ to the crossing equation with the same double discontinuity then they are related by the contact interactions in the AdS as follows
\begin{equation}\label{G1G2}
 G_1=G_2+c_1 G^{AdS}_{\phi^4}+c_2G^{AdS}_{(\partial\phi)^4}+c_2 G^{AdS}_{\phi^2(\partial^3\phi)^2}
\end{equation}
Furthermore, the authors showed\footnote{via explicit numerical computation} that for the four-point function of single trace scalar operator in Chern-Simons coupled fundamental scalar/fermion theories these $AdS_4$ contact terms do not contribute and hence the double discontinuity completely determines the four-point functions. 

 Consider the normalized three point functions of the operators $\mathcal{O}_i$($i=1,2,3)$ which are defined as follows
\begin{equation}\label{nor3}
C_{(123)}=\frac{\langle \mathcal{O}_1\mathcal{O}_2\mathcal{O}_3\rangle}{\sqrt{\langle\mathcal{O}_1\mathcal{O}_1\rangle\langle\mathcal{O}_2\mathcal{O}_2\rangle\langle\mathcal{O}_3\mathcal{O}_3\rangle}}.
\end{equation}
In \cite{Maldacena:2011jn, Maldacena:2012sf, Turiaci:2018dht}, it was noticed that the square of this normalized coefficients in the quasi-fermionic theories ($C^2_{s,qf}$) are related to that of a single free Majorana fermion ($C^2_{s,ff}$)  as follows
\begin{eqnarray}
 C^2_{s,qf}&=&\frac{1}{\tilde{N}}C^2_{s,ff}\label{QF3p}
\end{eqnarray}
where $\tilde{N}$ is related to the the rank of the gauge group $N$ and coupling $\lambda_{qf}$ by,
\begin{eqnarray}
 \tilde{N}= 2N \frac{\sin(\pi \lambda_{qf})}{\pi \lambda_{qf}}.
\end{eqnarray}
Note that the normalized coefficients of quasi-fermionic theory and free fermionic theory are proportional to each other as given in eq.(\ref{QF3p}). Hence, the double discontinuity of the scalar four point function in the free fermionic theory is same as that of the quasi-fermionic theories up to an overall factor which depends only on $N$ and $\lambda_{qf}$. 

On the other hand, the square of the normalized coefficients of the quasi-bosonic theories ($C^2_{s,qb}$) are related to the theory of a free real boson ($C^2_{s,fb}$)  as follows
\begin{eqnarray}
 C^2_{s,qb}&=&\frac{1}{\tilde{N}}C^2_{s,fb}~~~s>0,\label{QB3p}\\
 C^2_{0,qb}&=&\frac{1}{\tilde{N}}\frac{1}{(1+\tilde{\lambda}_{qb}^2)}C^2_{0,fb}=\frac{1}{\tilde{N}}C^2_{0,fb}-\frac1{\tilde N}\frac{\tilde{\lambda}_{qb}^2}{(1+\tilde{\lambda}_{qb}^2)}C^2_{0,fb}.\label{QB03p}
\end{eqnarray}
where $\tilde N$ and $\tilde{\lambda}$ are related to   $N$ and coupling $\lambda_{qb}$ as
\begin{eqnarray}
 \tilde{N}= 2N \frac{\sin(\pi \lambda_{qb})}{\pi \lambda_{qb}},\\
 \tilde{\lambda}_{qb}=\tan( \frac{\pi\lambda_{qb}}{2}).
\end{eqnarray}
Note that unlike the normalized coefficients of the quasi-fermionic theories, in the quasi-bosonic theories, the spin $s=0$ and $s\neq0$ coefficients given above  have different factors in front of their free bosonic counterparts. In order to account for the second term on the RHS of eq.(\ref{QB03p}) one needs to  add a conformal partial wave with spin-0 exchange which is given by the well known $\bar{D}$-function with the correct pre-factor \cite{Turiaci:2018dht}. We now proceed to employ this technique for the supersymmetric case.

\subsection{Double discontinuity and the supersymmetric correlators}
Here, we utilize the technique described above to compute the four-point correlators for spin-$0$ operators $J^b_0$ and $J^f_0$ in our supersymmetric theory. Since we are considering correlators of identical external operators\footnote{ Although we have all the three-point correlators required, we  do not compute mixed correlators such as $\langle J_0^bJ_0^bJ_0^fJ_0^f\rangle$ here, currently a free theory analogue for such correlators is not clear. We reserve this issue for future investigations. }, only even spin operators will contribute to the block expansion.

\subsubsection{\texorpdfstring{$\langle J_0^b(x_1)J_0^b(x_2)J_0^b(x_3)J_0^b(x_4)\rangle$}{<J0b J0b J0b J0b>} }
The four point function of the $J_0^b$ operators is expressed as follows
\begin{eqnarray}
\langle J_0^b(x_1)J_0^b(x_2)J_0^b(x_3)J_0^b(x_4)\rangle=disc+\frac{ 1}{x_{13}^2x_{24}^2} F(u,v).
\end{eqnarray}
Here, disc corresponds to the disconnected part given by
\begin{equation}
disc=\frac{1}{x_{12}^2x_{34}^2}+\frac{1}{x_{13}^2x_{24}^2}+\frac{1}{x_{14}^2x_{23}^2}
\end{equation}
while $F(u, v)$ is given by
\begin{align}
F(u, v) = \frac{1}{u}\sum_{k} C^2_{\mathcal{O}\mathcal{O}O_k}G_{\Delta_k, J_k}(u, v)\label{rcorr}
\end{align}
In order to determine the double discontinuity, and hence the four-point functions in the supersymmetric case using the method described above, we first obtain all the normalized three-point functions coefficients defined in  eq.(\ref{nor3}).
Utilizing the two and three point functions obtained in the previous section, the normalized coefficients for spins $s=0$ operators that contribute to the  $J_0^b$  four point function are given as follows\footnote{Note that $C^{BBB}_{s,susy}=\frac{\langle J_0^bJ_0^bJ_s^b\rangle}{\langle J_0^bJ_0^b\rangle\sqrt{\langle J_s^bJ_s^b\rangle}}$ and $C^{BBF}_{s,susy}=\frac{\langle J_0^bJ_0^bJ_s^f\rangle}{\langle J_0^bJ_0^b\rangle\sqrt{\langle J_s^fJ_s^f\rangle}}$. }
\begin{eqnarray}
 C^{2(BBB)}_{0,susy}&=&\frac{1}{\tilde N}\frac{(1-\tilde \lambda^2)^2}{(1+\tilde \lambda^2)^2}~C^{2}_{0,fb,}\label{CBB}\\
 C^{2(BBF)}_{0,susy}&=&-\frac{4}{\tilde N}\frac{\tilde \lambda^2}{(1+\tilde \lambda^2)^2}C^{2}_{0,fb}.\label{CBF}
\end{eqnarray}
On the other hand the normalized coefficients involving one of operators with  non-zero spin ($s>0$)  are given by
\begin{eqnarray}
C^{2(BBB)}_{s,susy}=\frac{1}{\tilde N (1+\tilde{\lambda}_{qb}^2)^2}C^{2}_{s,fb}~~~~s>0,~~~\text{even}\label{CBJB}\\\
C^{2(BBF)}_{s,susy}=\frac{\tilde{\lambda}_{qb}^4}{\tilde N(1+\tilde{\lambda}_{qb}^2)^2}C^{2}_{s,fb}~~~~s>0,~~~\text{even},\label{CBJF}
\end{eqnarray}
where $C^{2}_{s,fb}$ is the corresponding normalized coefficient of three point function for the free bosonic theory. The derivations for the above relations are provided in the Appendix \ref{NTH}. 

Since, the above relations were computed in momentum-space, they must be converted to position-space to make our results useful.\footnote{$C^2_s$ will denote the OPE coefficient in position space while $\tilde C^2_s$ denotes OPE coeffcient in position space. The relation between OPE coefficients in position and momentum space are given in \cite{Isono:2018rrb}.}
\begin{eqnarray}
 \tilde C^{2(BBB)}_{0,susy}&=&\frac{1}{\tilde N}\frac{(1-\tilde \lambda_{qb}^2)^2}{(1+\tilde \lambda_{qb}^2)^2}~\tilde C^{2}_{0,fb,}\label{xCBB}\\
 C^{2(BBF)}_{0,susy}&=&\frac{8}{\pi^2}\frac{\tilde \lambda_{qb}^2}{\tilde N(1+\tilde \lambda_{qb}^2)^2}\tilde C^{2}_{0,fb}.\label{xCBF}
\end{eqnarray}
Similarly, for $s>0$ are given by
\begin{eqnarray}
\tilde C^{2(BBB)}_{s,susy}=\frac{1}{\tilde N (1+\tilde\lambda_{qb}^2)^2}\tilde C^{2}_{s,fb}~~~~s>0,\label{xCBJB}\\\
\tilde C^{2(BBF)}_{s,susy}=\frac{\tilde\lambda_{qb}^4}{\tilde N(1+\tilde\lambda_{qb}^2)^2}\tilde C^{2}_{s,fb}~~~~s>0,\label{xCBJF}
\end{eqnarray}
Note that we may re-express both  the spin zero coefficients given by eq.(\ref{xCBB}) and eq.(\ref{xCBF}) as follows
\begin{eqnarray}
\tilde C^{2(BBB)}_{0,susy}&=&\frac{1}{\tilde N (1+\tilde{\lambda}_{qb}^2)^2}\tilde C^{2}_{0,fb}+\frac{\tilde{\lambda}_{qb}^4-2\tilde{\lambda}_{qb}^2}{\tilde N (1+\tilde\lambda_{qb}^2)^2}\tilde C^{2}_{0,fb},\label{CBB2}
\end{eqnarray} 
Observe that $\tilde C^{2(BBB)}_{s,susy}$ in eq.(\ref{xCBJB}) and the first term of $\tilde C^{2(BBB)}_{0,susy}$ in (\ref{CBB2}) have the same pre-factor. This  is similar to the case of the quasibosnic case given in eq.(\ref{QB3p}) and eq.(\ref{QB03p}) reviewed earlier. Consider, now, the double discontinuity of the conformal blocks
\begin{align}
\text{dDisc}[G_{\Delta, J}(1-z, 1-\bar z)] = \sin^2(\frac{\pi}{2}(\Delta-J-2\Delta_{\phi}))G_{\Delta, J}(1-z, 1-\bar z)
\end{align}
where $\Delta_{\phi}$ being the conformal dimension of the external operator. Notice that for $\Delta = 2\Delta_{\phi}+J+2m$, the double-discontinuity vanishes. Therefore, for the double-trace exchange, the double-discontinuity vanishes. That is why the OPE of single-trace operators are sufficient to construct a function that has a double-discontinutiy equal to the four-point correlator. However, notice that the single-trace exchange $J^{FF}_0$ with quantum numbers $(\Delta, J) = (2, 0)$ also vanish. Coincidently, the double-trace operator $[J^b_0, J^b_0]_{0, 0}$ also has the same quantum numbers.\footnote{$[\mathcal{O}, \mathcal{O}]_{n, l} = \mathcal{O}\Box^n \partial_{\mu_1}\partial_{\mu_2}\cdots\partial_{\mu_l}\mathcal{O}-\text{traces}$ where $\mathcal{O}$ is a single-trace operator.} By inspection, we can see that the function below has the right double-discontinuity
\begin{align}
 F(u,v) &= \frac{1+\tilde{\lambda}_{qb}^4}{\tilde N (1+\tilde{\lambda}_{qb}^2)^2} f_{fb}(u,v)-\frac{8}{\tilde N}\frac{2\tilde{\lambda}_{qb}^2}{\pi^{5/2}(1+\tilde{\lambda}_{qb}^2)^2}\big[\bar{D}_{11\frac{1}{2}\frac{1}{2}}(u,v)+\bar{D}_{11\frac{1}{2}\frac{1}{2}}(v,u)+\frac{1}{u}\bar{D}_{11\frac{1}{2}\frac{1}{2}}(\frac{1}{u},\frac{v}{u})\big]\notag\\ &+c_1 G^{AdS}_{\phi^4}+c_2G^{AdS}_{(\partial\phi)^4}+c_3 G^{AdS}_{\phi^2(\partial^3\phi)^2}\label{4ptB}
\end{align}
where, the function $f_{fb}(u,v)$ is the free bosonic part given by.\footnote{Note that we may have used two separate tree-level $\phi^3$ exchange Witten diagrams corresponding to $\Delta=1$ and $\Delta=2$ bulk exchange with arbitrary coefficients instead \cite{Dolan:2000ut}. But Witten diagrams themselves admitting an expansion in contact terms would compound the problem. The $\bar D$-functions, therefore, represents the choice with the least number of contact terms and the right double-discontinuity.} 
\begin{align}
f_{fb}(u, v) = 4\frac{1+u^{1/2}+v^{1/2}}{u^{1/2}v^{1/2}}\label{fb}
\end{align}
The contact terms are explicitly provided in \ref{ContactBB}. Note that $c_1$ contains contribution from both single-trace and double-trace operators which we have separated in the following equation as $a_1$ and $\tilde c_1$
\begin{align}
 F(u,v) &= \frac{1+\tilde{\lambda}_{qb}^4}{\tilde N (1+\tilde{\lambda}_{qb}^2)^2} f_{fb}(u,v)-\frac{8}{\tilde N}\frac{2\tilde{\lambda}_{qb}^2}{\pi^{5/2}(1+\tilde{\lambda}_{qb}^2)^2}\big[\bar{D}_{11\frac{1}{2}\frac{1}{2}}(u,v)+\bar{D}_{11\frac{1}{2}\frac{1}{2}}(v,u)+\frac{1}{u}\bar{D}_{11\frac{1}{2}\frac{1}{2}}(\frac{1}{u},\frac{v}{u})\big]\notag\\ &+a_1 \bar D_{1111}(u, v)+\tilde c_1 G^{AdS}_{\phi^4}+c_2G^{AdS}_{(\partial\phi)^4}+c_3 G^{AdS}_{\phi^2(\partial^3\phi)^2}\label{4ptB2}
\end{align}
To determine $a_1$ we take the OPE limit. In the OPE limit, the conformal blocks go like \footnote{OPE limit: $u \to 0, ~v \to 1, with (v-1)/u^{1/2}$ fixed}
\begin{align}
G_{\Delta, J}(u, v) \approx \frac{J!}{2^J (h-1)_J}u^{\Delta/2}C^{h-1}_{J}(\frac{v-1}{2\sqrt{u}})
\end{align} 
For $(\Delta, J) = (2, 0)$ i.e. for  $J^f_0$ exchange, we have $G_{2, 0}(u, v) \approx u$ in the OPE limit. Since, we are interested in the single-trace operator $J^f_0$, hence, we have
\begin{align}
F(u, v) = \tilde C^{2(BBF)}_{0, susy}\label{singF}
\end{align}
In the OPE limit, we have for $\phi^4$ contact term
\begin{align}
\bar D_{1111}(u, v) \approx 2
\end{align}
By only looking at the single-trace contributions we obtain
\begin{align}
&a_1 = \frac{\tilde C^{2(BBF)}_{0, susy}}{2}
\end{align}
Now, we focus our attention to double-trace operators. Coefficient $\tilde c_1$ can now be determined by looking at the double-trace trace operator $[J^b_0, J^b_0]_{0, 0}$. Since, $(\Delta, J) = (2, 0)$ for the double-trace is same as that of the single-trace operator $J^f_0$, we use the same method to obtain $\tilde c_1$

Hence, we have determined the first coefficient of the AdS contact terms. The results are a little cumbersome and we report it in the appendix \ref{apf}. We leave the explicit computation of these ope coefficients for future work.
\begin{align}
\bar c_1 = \frac{1}{2}\bigg([\tilde C^{2(FFF)}_{0, susy}]_{[O, O]_{0, 0}}-\frac{1}{\tilde N}\frac{4\tilde\lambda_{qb}^2}{(1+\tilde\lambda_{qb}^2)^2\pi^2}\tilde C^{2}_{0, fb}-\frac{1+\tilde{\lambda}_{qb}^4}{\tilde N (1+\tilde{\lambda}_{qb}^2)^2}[\tilde C^{2}_{0, fb}]_{[O, O]_{0, 0}}\bigg)
\end{align}
\subsubsection{\texorpdfstring{$\langle J_0^f(x_1)J_0^f(x_2)J_0^f(x_3)J_0^f(x_4)\rangle$}{<J0f J0f J0f J0f>} }
The four point function of $J_0^f$ is given by the following expression
\begin{equation}
\langle J_0^f(x_1)J_0^f(x_2)J_0^f(x_3)J_0^f(x_4)\rangle=disc+\frac{1}{x_{13}^4x_{24}^4}\mathcal{G}(u,v)
\end{equation}
where, $disc$ denotes the disconnected piece given by
\begin{equation}
disc=\frac{1}{x_{12}^4x_{34}^4}+\frac{1}{x_{13}^4x_{24}^4}+\frac{1}{x_{14}^4x_{23}^4}
\end{equation}
while $F(u, v)$ is given by
\begin{align}
\mathcal{G}(u, v) = \frac{1}{u}\sum_{k} C^2_{\mathcal{O}\mathcal{O}O_k}G_{\Delta_k, J_k}(u, v)\label{rcorrf}
\end{align} We now proceed to determine the four-point function $J_0^f$ utilizing the same technique.
 The normalized coefficients that are required in this case  are given by\footnote{Note that $C^{FFF}_{s,susy}=\frac{\langle J_0^fJ_0^f J_s^f\rangle}{\langle J_0^fJ_0^f\rangle\sqrt{\langle J_s^f J_s^f\rangle}}$ and $C^{FFB}_{s,susy}=\frac{\langle J_0^fJ_0^fJ_s^b\rangle}{\langle J_0^fJ_0^f\rangle\sqrt{\langle J_s^bJ_s^b\rangle}}$. }
\begin{eqnarray}
C^{2(FFF)}_{s,susy}=\frac{\tilde{\lambda}_{qf}^4}{\tilde N (1+\tilde{\lambda}_{qf}^2)^2}C^{2}_{s,ff},\label{CFJF}\\
C^{2(FFB)}_{s,susy}=\frac{1}{\tilde N (1+\tilde{\lambda}_{qf}^2)^2}C^{2}_{s,ff},\label{CFJB}
\end{eqnarray}
where $C^{2}_{s,ff}$ is the normalized three point functions for free fermionic theory. Unlike the previous section, changing the OPE coefficients to position-space is redundant here as both sides of the eqaution change by the same factor. Once again these relations are derived in Appendix \ref{NTH}. Note that the three point functions of the spin-0 exchanges given by $C^{2(FFF)}_{0,susy}$ and $C^{2(FFB)}_{0,susy}$ are contact terms in this case which, therefore, may be set to zero. This implies that the above relation is trivially satisfied for the spin $s=0$ case as the free fermionic coeffcient $C^{2}_{0,ff}=0$. Hence, both the $s=0$ and $s\neq0$ coefficients in this case come with the same pre-factor. This implies that the function which has the correct double discontinuity is given by
\begin{equation}\label{freef}
\mathcal{G}(u,v)=\frac{1+\tilde{\lambda}_{qf}^4}{\tilde N (1+\tilde{\lambda}_{qf}^2)^2} f_{ff}(u,v)+\bar c_1 G^{AdS}_{\phi^4}+\bar c_2G^{AdS}_{(\partial\phi)^4}+\bar c_3 G^{AdS}_{\phi^2(\partial^3\phi)^2},
\end{equation}
where $f_{ff}(u,v)$ is the free fermionic part given by
\begin{align}
\label{DBs2}
f_{ff}(u, v) = \frac{1+u^{5/2}+v^{5/2}-u^{3/2}(1+v)-v^{3/2}(1+u)-u-v}{u^{3/2}v^{3/2}}
\end{align}
However, in this case, the separation of the double-trace and the single trace contributions is redundant as there are no single-trace operators with $(\Delta, J) = (4, 0)$. After a faithful implementation of the methodology of the previous section, we obtain 

\begin{align}
\bar c_1 = \frac{3\pi^{1/2}}{8P^{(2)}_1(0, 0)}\bigg([\tilde C^{2(FFF)}_{0, susy}]_{[O, O]_{0, 0}}-\frac{1+\tilde{\lambda}^4}{\tilde N (1+\tilde{\lambda}^2)^2}[\tilde C^{2}_{0, ff}]_{[O, O]_{0, 0}}\bigg)
\end{align}
This completes our analysis of the determination of the four point functions of $J_0^b$ and $J_0^f$ operators in the Chern Simons matter theories in the supersymmetric  scenario. The contact terms are explicitly provided in \ref{ContactFF}. The results are a little cumbersome and we report it in the appendix \ref{apf}. We leave the explicit computation of these ope coefficients for future work.


\section{Summary and Discussion}\label{discus}
In this article, we have focused our attention on the ${\cal N}=2$ $U(N)$ Chern Simons theory coupled with 
a single fundamental chiral multiplet in the 't Hooft large $N$ limit and presented the computations for the exact 
2 and 3-point functions for the scalar supermultiplet. The result are invariant the duality transformation \eqref{dualityst} 
and can be seen as an independent confirmation of the duality. 
For the case of 4-point function, though we are not able to perform the direct computation for the full scalar supermultiplet, 
we are able to use the a combination of techniques from conformal bootstrap, factorization of 3-point functions via double 
trace interactions along with the self duality of our theory to write down two of the component of the full $J_0$ 4-point function. 
Though this leave room for 3 undetermined coefficient, we could formally related these to 3 point function coefficients involving 
specific double trace operators. We plan to report on these in near future. 

Though we have focused on the ${\cal N}=2$ theory in this paper, the approach used to compute the four point function is 
can be straightforwardly applied to the one parameter deformed ${\cal N}=1$ theory. These differ from our ${\cal N}=2$ 
theory only via a double trace term in ${\cal N}=1$ superspace\footnote{$\delta S = \frac{\pi w}{\kappa}\int d^3xd^2\theta (\bar\Phi\Phi)^2$}. 
The 2 and 3-point functions of the two theories can thus be related via the double trace type factorization also used in this paper. 

The approach used in this paper, following \cite{Turiaci:2018dht}, to compute the $J_0^b$ and $J_0^f$ 4-point functions relies 
crucially on the fact that the double discontinuity of the 4-point function in the interacting theory is almost the same as that of the 
free theory. We could thus write down the full interacting 4-point function in term of the free 4-point function. For the case of 
mixed 4-point functions, e.g. $\la J_0^b J_0^b J_0^f J_0^f \ra$, this approach is not directly useful since a free theory analogue 
of such mixed correlator is not available since bosons and fermions decouple from each other the mixed 4 point correlators vanish 
in $\lambda \rightarrow 0$ limit. One approach that might be useful 
in this regard is to first study the single trace OPE coefficients in the ${\cal N}=1$ deformed theory (for general $w$) in $\lambda=0$ limit.  
We expect this limit to be significantly simpler then ${\cal N}=2$ theory and one can compute not only the exact
2, 3 point functions (see e.g. \cite{Aharony:2019mbc}) but perhaps even the exact 4-point function (we expect it to be 
non vanishing for $w \neq 0$) of $J_0$ operators in this limit since the only interaction term present is a double trace 
term. If this is indeed turn out to be the case, one can compare the double discontinuity of mixed 
$J_0^b, J_0^f$ correlators in ${\cal N}=2$ theory with this limit and see these are closely related in a similar way as 
in \cite{Turiaci:2018dht} and in this paper for the identical scalar 4-point function. 

As we have noticed in this paper, the coefficients $\{c_i\}$ can be determined in term of the normalized 
3-point function coefficients of specific double trace operator. An interesting property of the $AdS_4$ contact Witten diagrams is 
that their series expansions contain $Log$ terms. This implies that the coefficients $\{c_i\}$ not only contribute to the OPE coefficients 
of double trace terms but also to their leading anomalous dimensions as well, but in a coordinated way. The absence of 
$Log$ term in the free 4-point function along with the vanishing of these coefficients for quasi-bosonic and quasi-fermionic 
theories \cite{Turiaci:2018dht} thus means that these double trace operators in the leading large-N order do not receive corrections to there 
anomalous dimensions these theories. Whether this is also the case in the supersymmetric theory studied in this paper as well 
requires the computation of anomalous dimensions of these double trace operators which we leave for future investigation.

\section*{Acknowledgements}
We would like to thank O Aharony, A Gadde, S Minwalla, Naveen Prabhakar and A. Sharon for fruitful discussions. 
TS would like to thanks Antal Jevicki for many enjoyable discussions on Chern Simons matter theories and related topics. 
SJ and KI also thank the organizers of the Batsheva de Rothschild Seminar on Avant-garde methods for quantum field theory and 
gravity, for hospitality. The work of KI was supported in part by a center of excellence supported by the Israel Science 
Foundation (grant number 1989/14), the US-Israel bi-national fund (BSF) grant number 2012383 and the Germany 
Israel bi-national fund GIF grant number I-244-303.7-2013 at Tel Aviv University and BSF grant number 2014707 
at Ben Gurion University. Research of SJ and VM is supported by Ramanujan Fellowship. 
Reserach work of TS is supported by Simons Foundation Grant Award 509116 and 
Ramanujan Fellowship. AM would like to acknowledge the support of CSIR-UGC (JRF) 
fellowship (09/936(0212)/2019-EMR-I). PN acknowledges support from the College of Arts and Sciences of the 
University of Kentucky. . Finally SJ, VM, AM would like to acknowledge 
our debt to the steady support of the people of India for research in the basic sciences.


\appendix

\section{Notations and Conventions}\label{NandC}
\be
\begin{split}
\textrm{Metric} & : \eta_{\mu\nu} = \textrm{diag}(-1,1,1) \\
\textrm{Gamma Matrices} & : (\gamma^\mu)_\A^{~\B} = (\sigma_2, -i\sigma_1, i\sigma_3)_\A^{~\B}  \quad 
                        \Rightarrow \{ \gamma^\mu, \gamma^\nu \} = -2\eta^{\mu\nu} I_2 \\
\textrm{Charge Conjugation} &: C_{\A\B} = -C_{\B\A} = \begin{pmatrix} 0 & -i \\ i & 0 \end{pmatrix} = -C^{\A\B} = C^{\B\A} \\
\textrm{Raising-Lowering} & : \psi^\A = C^{\A\B} \psi_\B \quad ; \quad \psi_\A = -C_{\A\B} \psi^\B = \psi^\beta C_{\beta\alpha}\\ 
                & \Rightarrow \psi^+ = i\psi_- \quad ; \quad \psi^- = -i \psi_+ \\
\textrm{Vector $\leftrightarrow$ Bi-spinor} & : p_{\A\B} = p_\mu (\gamma^\mu)_{\A\B} 
             = \begin{pmatrix} p_0+p_1 & p_3 \\ p_3 & p_0-p_1 \end{pmatrix} 
             = \begin{pmatrix} p_+ & p_3 \\ p_3 & -p_- \end{pmatrix}\\
\textrm{Squared Grassmann variables } & : \theta^2 = \frac12 \theta^\alpha \theta_\alpha, \ d^2\theta = \frac 12 d\theta^\alpha d\theta_\alpha \\
\textrm{Superspace integrals } & : \int d\theta=0, \ \int d\theta \ \theta = 1\\
                        & \int d^2\theta \ \theta^2 = -1, \ \int d^2\theta \ \theta^\alpha \theta^\beta = C^{\alpha\beta}\\
\textrm{Grassmann $\delta$-function } & : \delta^2(\theta) = -\theta^2\\
\textrm{Superfields } & : \Phi = \phi+\theta \psi-\theta^2 F, \ \bar \Phi = \bar \phi+\theta \bar \psi-\theta^2 \bar F\\
                &  \bar\Phi\Phi = \bar\phi\phi+\theta^\alpha\left(\bar\phi \psi_\alpha + \bar \psi_\alpha \phi \right) - \theta^2 \left(\bar F \phi +\bar \phi F+ \bar \psi \psi \right)
\end{split}
\ee

\section{Component 3 point functions}\label{c3ptfnWI}
In this appendix, we write down the component 3 functions abstractly in term of the functions $\{ A_i \}$ appearing in 
form of full superspace 3 point function \eqref{J03ptresult} determined by supersymmetric Ward identity. 
\be
\begin{split}
\la J^b_0(p) J^b_0(-p-s) J^b_0(s) \ra & = 2 A_1 \\
\la J^f_0(p) J^f_0(-p-s) J^f_0(s) \ra & = 2 (A_3 p_3^2 + s_3 (-A_4 p_3 - A_5 p_3 + A_2 s_3)) \\
\la J^b_0(p) J^f_0(-p-s) J^b_0(s) \ra & = 2 (A_2 + A_3 + A_4 + A_5) \\
\la J^f_0(p) J^b_0(-p-s) J^f_0(s) \ra  & = \frac{2}{9} (9 A_6 + (p_3 - s_3) (3 A_4 - 3 A_5 + A_1 p_3 - A_1 s_3)) \\
\la \Psi_+(p) J^b_0(-p-s) \Psi_-(s) \ra & = -\frac{2}{3} (3 A_5+A_1 (-p_3+s_3)) \\
\la \Psi_+(p) J^b_0(-p-s) \Psi_-(s) \ra & = -\frac{2}{9} (-9 A_6 + p_3 (-3 (3 A_3+A_4+2 A_5) + 2 A_1 p_3) \\ 
                                                            &   \qquad + (9 A_2+3 A_4+6 A_5+5 A_1 p_3) s_3+2 A_1 s_3^2)
\end{split}
\ee

\section{Normalized higher spin three point functions}\label{NTH}
In this section, we aim to derive the normalized coefficients of the three point functions involving two $J_{0}$ operators and one higher spin operator $J_{s}$ ($s>1$) leading to eq.(\ref{CBJB}), (\ref{CBJF}), (\ref{CFJF}) and (\ref{CFJB}). The classical action for ${\cal N}=2$ supersymmetric matter coupled to Chern Simons gauge field is given by
 \begin{align}
& S^{\mathcal{T}}_{k, N} = \frac{ik}{4\pi}S_{CS}(A)+S_{\mathcal{B}_{k, N}}(A, \phi)+S_{\mathcal{F}_{k, N}}(A, \psi)+S_{bf}(\varphi, \psi)\notag
\end{align}
\begin{align}
&S_{CS}(A) = \int d^3x ~\epsilon_{\mu\nu\rho}\text{Tr}(A^{\mu}\partial^{\nu}A^{\rho}-\frac{2i}{3}A^{\mu}A^{\nu}A^{\rho})\notag\\
&S_{\mathcal{B}_{k, N}}(A, \phi) = \frac{ik}{4\pi}S_{CS}(A) + \mathcal{D}_{\mu}\bar\varphi\mathcal{D}^{\mu}\varphi\notag\\
&S_{\mathcal{F}_{k, N}}(A, \psi) = \frac{ik}{4\pi}S_{CS}(A) - i\bar\psi\gamma^{\mu}\mathcal{D}_{\mu}\psi\\
& S_{bf}(\varphi, \psi) = \int d^3 x [-\frac{4\pi i}{k}(\bar\varphi\varphi)(\bar\psi\psi)+\frac{4\pi^2}{k^2}(\bar\varphi\varphi)^3-\frac{2\pi i}{k}(\bar\psi\varphi)(\bar\varphi\psi)]\notag
\end{align}
Note that we will be using the following notation for correlation functions in the superymmetric and the non-supersymmetric theories
\begin{align}
&\langle\cdots\rangle_{\mathcal{T}_{k, N}} ~~~~~~~\text{Correlator in SUSY theory}\notag\\
&\langle\cdots\rangle_{\mathcal{B}_{k, N}} ~~~~~~~\text{Correlator in regular bosonic theory}\notag\\
&\langle\cdots\rangle_{\mathcal{F}_{k, N}} ~~~~~~~\text{Correlator in regular fermionic theory}\notag\\
&\langle\cdots\rangle_{fb} ~~~~~~~~~~\text{Correlator in free bosonic theory}\notag\\
&\langle\cdots\rangle_{ff} ~~~~~~~~~~\text{Correlator in free fermionic theory}\notag\\
\end{align}
We begin with the derivation of relation between the normalized three point functions of the $\langle J_s^b J_0^b J_0^b\rangle$ in supersymmetric theory and the corresponding three point function in the free bosonic theory. We will employ the path integral technique and utilize some of the relations derived in \cite{Gur-Ari:2015pca}. The correlation function $\langle J_s^b J_0^b J_0^b\rangle_{\mathcal{T}_{k, N}}$ in the path integral representation may be expressed as follows
\begin{align}
&\langle J_s^b(x_a) J_0^b(x_b) J_0^b(x_c) \rangle_{\mathcal{T}_{k, N}}\notag\\ 
&= \int \mathcal{D}\phi~ J_s^b(x_a) J_0^b(x_b) J_0^b(x_c) ~e^{-S^{\mathcal{B}}_{k, N}-S^{\mathcal{F}}_{k, N}}(1-S_{bf}+\frac{1}{2!}S^2_{bf}-\frac{1}{3!}S^3_{bf}+\cdots)\notag\\
& = \langle J_s^b(x_a) J_0^b(x_b) J_0^b(x_c) \rangle_{\mathcal{B}_{k, N}}\notag\\&-\frac{4\pi}{k}\int_1 \langle  J_s^b(x_a) J_0^b(x_b) J_0^b(x_c)J_0^b(x_1) \rangle_{\mathcal{B}_{k, N}}\langle  J_0^f(x_1)\rangle_{\mathcal{F}_{k, N}}\notag\\&+\frac{1}{2!}(\frac{4\pi}{k})^2\int_{12}\langle J_s^b(x_a) J_0^b(x_b) J_0^b(x_c)J_0^b(x_1)J_0^b(x_2) \rangle_{\mathcal{B}_{k, N}}\langle J_0^f(x_1)J_0^f(x_2)\rangle_{\mathcal{F}_{k, N}}\notag\\&-\frac{1}{3!}(\frac{4\pi}{k})^3\int_{123} \langle J_s^b(x_a) J_0^b(x_b) J_0^b(x_c)J_0^b(x_1)J_0^b(x_2)J_0^b(x_3) \rangle_{\mathcal{B}_{k, N}}\langle  J_0^f(x_1)J_0^f(x_2)J_0^f(x_3)\rangle_{\mathcal{F}_{k, N}}+\cdots
\end{align}
In the first line, we have implemented factorization of planar correlators through the multi-trace interactions of the SUSY theory by splitting the action into non-SUSY components\cite{Gur-Ari:2015pca}. Now, by usual arguments of connectedness and by Wick's theorem, 
\begin{align}
&\langle J_s^b(x_a) J_0^b(x_b) J_0^b(x_c) \rangle_{\mathcal{T}_{k, N}} \notag\\
&= \langle J_s^b(x_a) J_0^b(x_b) J_0^b(x_c) \rangle_{\mathcal{B}_{k, N}}+2(\frac{4\pi}{k})^2\int_{12} \langle J_0^b J_0^b(x_1)\rangle_{\mathcal{B}_{k, N}}\langle J^b_s J_0^bJ_0^b(x_2) \rangle_{\mathcal{B}_{k, N}}\langle J_0^f(x_1)J_0^f(x_2)\rangle_{\mathcal{F}_{k, N}}+\cdots\notag\\
&=\langle J_s^b(x_a) J_0^b(x_b) J_0^b(x_c) \rangle_{\mathcal{B}_{k, N}}(1+\sum_{n = 1}^{\infty}(n+1)[(\frac{4\pi}{k})^2\langle J_0^bJ_0^b\rangle_{\mathcal{B}_{k, N}}\langle J_0^fJ_0^f\rangle_{\mathcal{F}_{k, N}}]^n)\notag\\
&=\langle J_s^b(x_a) J_0^b(x_b) J_0^b(x_c) \rangle_{\mathcal{B}_{k, N}}\frac{1}{(1-(\frac{4\pi}{k})^2\langle J_0^bJ_0^b\rangle_{\mathcal{B}_{k, N}}\langle J_0^fJ_0^f\rangle_{\mathcal{F}_{k, N}})^2}\label{gur}
\end{align}
The various integrations and the position labels in the above are implicit. The implicit notation and the momentum space representation are look identical and therefore, the implicit format may be viewed in momentum-space representation. In  eq.(20) of \cite{Gur-Ari:2015pca}, an interesting relation was derived between the two point function which is given as
\begin{equation}
 \langle J_0^b J_0^f \rangle_{\mathcal{T}_{k, N}}=-\frac{k}{4 \pi}\sum_{n=1}^{\infty} (\frac{4 \pi}{k})^2\langle J_0^bJ_0^b\rangle_{\mathcal{B}_{k, N}}\langle J_0^fJ_0^f\rangle_{\mathcal{F}_{k, N}}
\end{equation}
Hence eq.(\ref{gur}) simplifies upon utilizing the above relation as
\begin{align}
\langle J_s^b(x_a) J_0^b(x_b) J_0^b(x_c) \rangle_{\mathcal{T}_{k, N}} &=\langle J_s^b(x_a) J_0^b(x_b) J_0^b(x_c) \rangle_{\mathcal{B}_{k, N}}[1-\frac{4\pi}{k}\langle J_0^b J_0^f \rangle_{\mathcal{T}_{k, N}}]^2 \notag \\
&=\langle J_s^b(x_a) J_0^b(x_b) J_0^b(x_c) \rangle_{\mathcal{B}_{k, N}}\frac{1}{(1+\tilde\lambda^2)^2}\label{obbb1}
\end{align}
Note that in order to arrive at the last line in the above relation we have utilized the explicit form of the two point function $\langle J_0^b J_0^f \rangle_{\mathcal{T}_{k, N}}$ which we had derived earlier.
The two point functions in the supersymmetric theory are related to the regular bosonic theory as follows \cite{Gur-Ari:2015pca}
\begin{eqnarray}
 \langle J_s^b J_s^b \rangle_{\mathcal{T}_{k, N}} & =& \langle J_s^b J_s^b\rangle_{\mathcal{B}_{k, N}}\label{o0}\\
 \langle J_0^b J_0^b \rangle_{\mathcal{T}_{k, N}} 
&=&\langle J_0^b J_0^b\rangle_{\mathcal{B}_{k, N}}[1-\frac{4\pi}{k}\langle J_0^b J_0^f \rangle_{\mathcal{T}_{k, N}} ]\label{os}
\end{eqnarray}
From eq.(\ref{obbb1}), eq.(\ref{o0}) and eq.(\ref{os}) we obtain the normalized coefficients as
\begin{equation}
 \frac{\langle J_s^b J_0^b J_0^b \rangle_{\mathcal{T}_{k_e, N_e}}}{\langle J_0^b J_0^b\rangle_{\mathcal{T}_{k_e, N_e}}\sqrt{\langle J_s^b J_s^b\rangle_{\mathcal{T}_{k_e, N_e}}}} =\frac{1}{\sqrt{\tilde N}(1+\tilde\lambda^2)}\frac{\langle J_s^b J_0^b J_0^b \rangle_{fb}}{\langle J_0^b J_0^b\rangle_{fb}\sqrt{\langle J_s^b J_s^b\rangle_{fb}}}\label{jbbb}
\end{equation}
where, we have made use of the relation between the normalized coefficients of the regular bosonic theory and the free bosonic theory  derived in \cite{Maldacena:2011jn,Maldacena:2012sf} 
\begin{equation}
 \frac{\langle J_s^b J_0^b J_0^b \rangle_{\mathcal{B}_{k_e, N_e}}}{\langle J_0^b J_0^b\rangle_{\mathcal{B}_{k_e, N_e}}\sqrt{\langle J_s^b J_s^b\rangle_{\mathcal{B}_{k_e, N_e}}}}=\frac{1}{\sqrt{\tilde N}}\frac{\langle J_s^b J_0^b J_0^b \rangle_{fb}}{\langle J_0^b J_0^b\rangle_{fb}\sqrt{\langle J_s^b J_s^b\rangle_{fb}}}
\end{equation}
Thus, from eq.(\ref{jbbb}) we have obtained the required relation
\begin{equation}
 C^{2(BBB)}_{s,susy}=\frac{1}{\tilde N (1+\tilde{\lambda}_{qb}^2)^2}C^{2}_{s,fb}
\end{equation}
where
\begin{equation}
 \tilde{\lambda}_{qb}=\tilde{\lambda}=\tan(\frac{\pi \lambda}{2})
\end{equation}

A similar calculation allows one to write 
\begin{align}
\langle J_s^f(x_a) J_0^f(x_b) J_0^f(x_c) \rangle_{\mathcal{T}_{k, N}} =\langle J_s^f(x_a) J_0^f(x_b) J_0^f(x_c) \rangle_{\mathcal{F}_{k, N}}[1-\frac{4\pi}{k}\langle J_0^b J_0^f \rangle_{\mathcal{T}_{k, N}}]^2
\end{align}
 Upon using the two point functions leads to the normalized coefficient
\begin{equation}
 \frac{\langle J_s^f J_0^f J_0^f \rangle_{\mathcal{T}_{k_e, N_e}}}{\langle J_0^f J_0^f\rangle_{\mathcal{T}_{k_e, N_e}}\sqrt{\langle J_s^f J_s^f\rangle_{\mathcal{T}_{k_e, N_e}}}}=\frac{1}{\sqrt{\tilde N}(1+\tilde\lambda^2)}\frac{\langle J_s^f J_0^f J_0^f \rangle_{ff}}{\langle J_0^f J_0^f\rangle_{ff}\sqrt{\langle J_s^f J_s^f\rangle_{ff}}}\label{ffFF}
\end{equation}
Hence we obtain the required coefficient as
\begin{equation}
 C^{2(FFF)}_{s,susy}=\frac{1}{\tilde N (1+\tilde{\lambda}_{qb}^2)^2}C^{2}_{s,ff}=\frac{\tilde{\lambda}_{qf}^4}{\tilde N (1+\tilde{\lambda}_{qf}^2)^2}C^{2}_{s,ff}
\end{equation}
where
\begin{equation}
 \tilde{\lambda}_{qf}=\frac{1}{\tilde{\lambda}}=\cot(\frac{\pi \lambda}{2})
\end{equation}
In order to derive the other two coefficients in  eq.(\ref{CBJF}) and eq.(\ref{CFJB}), we make use of the Giveon-Kutasov duality which relates the magnetic theory to the electric counterpart through the following exchanges\cite{Giveon:2008zn,Benini:2011mf,Gur-Ari:2015pca}
\begin{align}
J_0^f \leftrightarrow -J_0^f ~~ J_0^b \leftrightarrow -J_0^b ~~ J_s^f \leftrightarrow (-1)^s J_s^b\label{37}
\end{align}
 to write the following
\begin{align}
\langle J_s^f J_0^f J_0^f \rangle_{\mathcal{T}_{k_e, N_e}} = (-1)^s\langle J_s^b J_0^f J_0^f \rangle_{\mathcal{T}_{k_m, N_m}}\\
\langle J_s^b J_0^b J_0^b \rangle_{\mathcal{T}_{k_e, N_e}} = (-1)^s\langle J_s^f J_0^b J_0^b \rangle_{\mathcal{T}_{k_m, N_m}}
\end{align}
where now the labels $(k_e, N_e)$ and $(k_m, N_m)$ represent the electric $(e)$ and the magnetic $(m)$ theories, respectively. Now, by the normalization convention of (7), we can rewrite the following, 
 \begin{align}
\frac{\langle J_s^b J_0^f J_0^f \rangle_{\mathcal{T}_{k_e, N_e}}}{\langle J_0^f J_0^f\rangle_{{\cal T}_{k_e,N_e}} \sqrt{\langle J_0^f J_0^f\rangle}_{\mathcal{T}_{k_e, N_e}}}&=(-1)^s\frac{\langle J_s^f J_0^f J_0^f \rangle_{\mathcal{T}_{k_m, N_m}}}{ \langle J_0^f J_0^f\rangle_{{\cal T}_{k_m,N_m}}\sqrt{\langle J_s^f J_s^f\rangle_{{\cal T}_{k_m,N_m}}}}\\
&=\frac{1}{(1+\tilde\lambda_m^2)}\frac{\langle J_s^f J_0^f J_0^f \rangle_{\mathcal{F}_{k_m, N_m}}}{\langle J_0^f J_0^f\rangle_{\mathcal{F}_{k_m, N_m}}\sqrt{\langle J_s^f J_s^f\rangle_{\mathcal{F}_{k_m, N_m}}}}\notag\\
&=\frac{\tilde\lambda_e^2}{\sqrt{\tilde N_e}(1+\tilde\lambda_e^2)}\frac{\langle J_s^f J_0^f J_0^f \rangle_{ff}}{\langle J_0^f J_0^f\rangle_{ff}\sqrt{\langle J_s^f J_s^f\rangle_{ff}}}\label{ffjb}
\end{align}
The first equality uses the duality described in eq.(\ref{37}). The second equality makes use of the relation derived in eq.(\ref{ffFF}). The third equality utilizes the results of Maldacena and Zhiboedov derived in \cite{Maldacena:2011jn,Maldacena:2012sf} and the relation between the parameters of the magnetic and electric theory namely $\tilde N_m=\tilde N_e$ and $\tilde\lambda_m=\frac{1}{\tilde\lambda_e}$. This leads to the following expression for the normalized coefficient 
\begin{equation}
 C^{2(FFB)}_{s,susy}=\frac{\tilde\lambda^4}{\tilde N (1+\tilde\lambda^2)^2}C^{2}_{s,ff}=\frac{1}{\tilde N(1+\tilde{\lambda}_{qf}^2)^2}C^{2}_{s,ff}
\end{equation}

Similar computations allow us to write
\begin{align}
\frac{\langle J_s^f J_0^b J_0^b \rangle_{\mathcal{T}_{k_e, N_e}}}{\langle J_0^b J_0^b\rangle_{{\cal T}_{k_e,N_e}} \sqrt{\langle J_s^f J_s^f \rangle_{\mathcal{T}_{k_e, N_e}}}}&=(-1)^s\frac{\langle J^b_s J_0^b J_0^b \rangle_{\mathcal{T}_{k_m, N_m}}}{ \langle J_0^b J_0^b\rangle_{{\cal T}_{k_m,N_m}}\sqrt{\langle J_s^b J_s^b\rangle_{{\cal T}_{k_m,N_m}}}}\\
& =\frac{1}{1+\tilde\lambda_m^2}\frac{\langle J_s^b J_0^b J_0^b \rangle_{\mathcal{B}_{k_m, N_m}}}{\langle J_0^b J_0^b\rangle_{\mathcal{B}_{k_m, N_m}}\sqrt{\langle J_s^b J_s^b\rangle_{\mathcal{B}_{k_m, N_m}}}}\label{bbjb}\\
&=\frac{1}{\sqrt{\tilde N}_m(1+\tilde\lambda_m^2)}\frac{\langle J_s^b J_0^b J_0^b \rangle_{fb}}{\langle J_0^b J_0^b\rangle_{fb}\sqrt{\langle J_s^b J_s^b\rangle_{fb}}}\notag\\
\end{align}
This leads to the normalized coefficient $C^{2(BBF)}_{s,susy}$ upon using $\tilde N_m=\tilde N_e$ and $\tilde\lambda_m=\frac{1}{\tilde\lambda_e}$
\begin{equation}
 C^{2(BBF)}_{s,susy}=\frac{\tilde\lambda^4}{\tilde N(1+\tilde\lambda^2)^2}C^{2}_{s,fb}=(\frac{\tilde{\lambda}_{qb}^4}{\tilde N(1+\tilde{\lambda}_{qb}^2})^2C^{2}_{s,fb}
\end{equation}

\section{Comments on direct computation of \texorpdfstring{${\cal J}^{(0)}$}{J0} 4 point function}\label{4ptcomp}
\begin{figure}[h]
\includegraphics[width=1\textwidth]{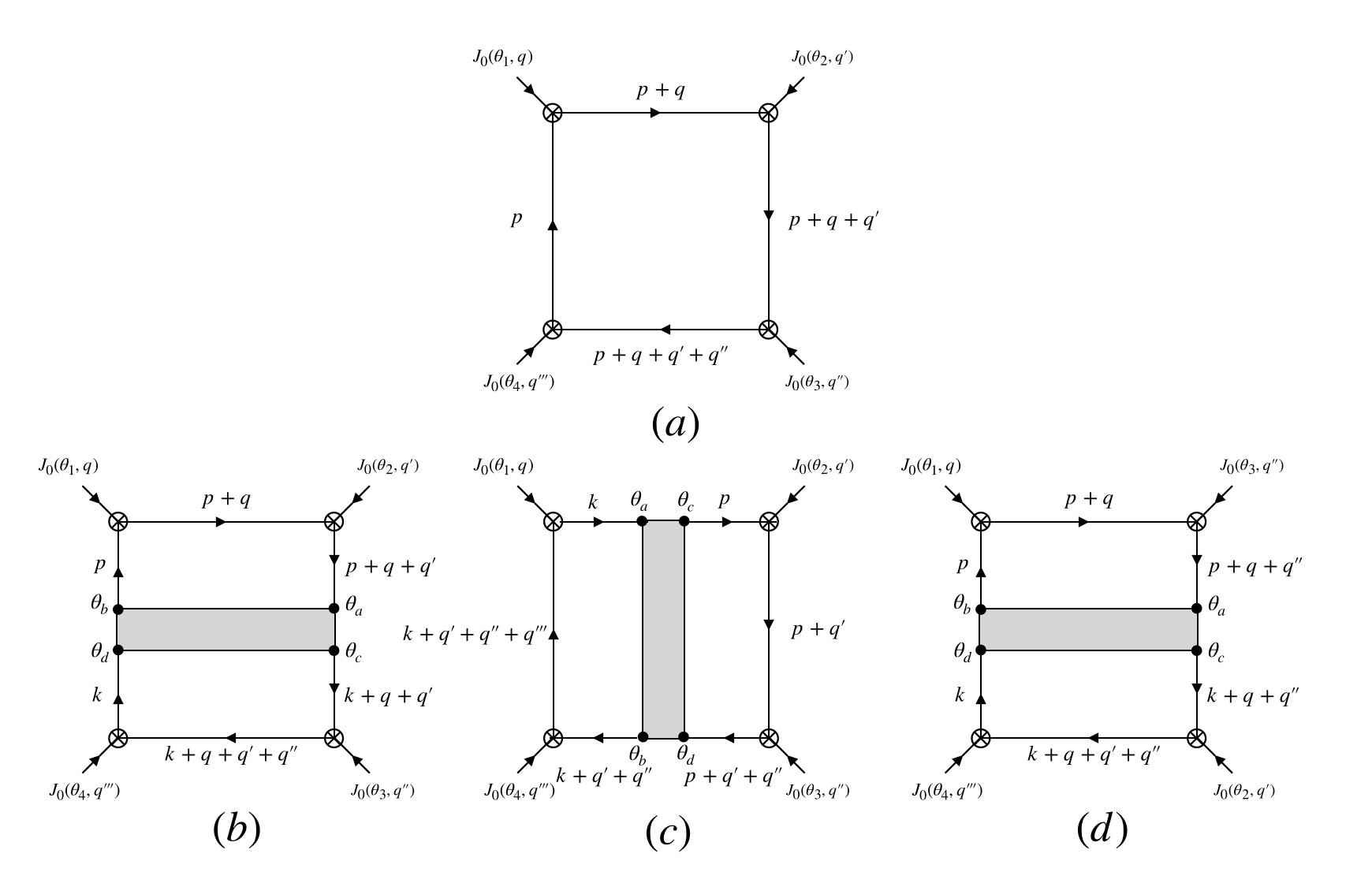}
\caption{The contributing diagrams for the four point function of currents. The first diagram is diagram type (a). 
The grey blob in (b), (c) , (d) represents the all loop four point correlator. The remaining diagrams are obtained 
by permutations of the external operators. \label{4ptdiags} }
\end{figure}

In this appendix, we will describe the relevant diagrams, and corresponding integrals, constructed using the 
exact 4 point vertex which contribute to the full ${\cal J}^{(0)}$ four point function. Figure \ref{4ptdiags} 
show all the relevant diagrams. 

For diagram one in \ref{4ptdiags}, Note that the exact vertex \eqref{J0vert} is a function of two internal grassmann variables. 
The propagators in fig \eqref{4ptdiags} that emanate from/to the crossed vertices connect these internal Grassmann variables. 
So one has additional integration of 8 two-component internal Grassmann variables, these are not shown in the figure.  
In the exact three vertex (crossed), for the internal theta arguments, ${\mathcal V}_3(\theta_1,\theta_1',\theta_1'', q, p)$ 
the primed theta is on the direction of the leaving momenta, and the double primed momenta are on the direction of the 
incoming momenta. The insertion momentum is the first momentum argument, and the incoming momentum is the second 
momentum argument.
In fig \eqref{4ptdiags} the diagram a is given by
\begin{align}
&V^{(A)}(q,q',q'',\theta_1,\theta_2,\theta_3,\theta_4)\nonumber\\
&= N \int \frac{d^3p}{(2\pi)^3} d^2\theta_1'd^2\theta_1''d^2\theta_2'd^2\theta_1'' d^2\theta_3' d^2\theta_3''d^2\theta_4' d^2\theta_4''\nonumber\\
&\biggl(P(\theta_1',\theta_4'',p+q) P(\theta_4',\theta_3'',p-q'-q'') P(\theta_3',\theta_2'',p-q')P(\theta_2',\theta_1'',p)\nonumber\\
&{\mathcal V}_3(\theta_1,\theta_1',\theta_1'', q, p){\mathcal V}_3(\theta_2,\theta_2',\theta_2'',q',p-q'){\mathcal V}_3(\theta_3,\theta_3',\theta_3'',q'',p-q'-q''){\mathcal V}_3(\theta_4,\theta_4',\theta_4'',-q-q'-q'',p+q)\biggr)\label{VA}
\end{align}
There are a total of 6 additional diagrams due to permutations of the operators. and the interaction part is given by
\begin{align}
    & V^{(B)}_4(q,q',q'',\theta_1,\theta_2,\theta_3,\theta_4)\nonumber\\
    &=N^2\int \frac{d^3p}{(2\pi)^3}\frac{d^3k}{(2\pi)^3} d^2\theta_a d^2\theta_b d^2\theta_c d^2\theta_d d^2\theta_1'd^2\theta_1''d^2\theta_2'd^2\theta_1'' d^2\theta_3' d^2\theta_3''d^2\theta_4' d^2\theta_4''\nonumber\\
    &\biggl(P(\theta_1',\theta_4'',p+q) P(\theta_4',\theta_a,p-q'-q'') P(\theta_c,\theta_3'',k-q'-q'')P(\theta_3',\theta_2'',k-q')P(\theta_2',\theta_d,k)P(\theta_b,\theta_1'',p)\nonumber\\
    &{\mathcal V}_3(\theta_1,\theta_1',\theta_1'', q, p){\mathcal V}_3(\theta_2,\theta_2',\theta_2'',q',k-q'){\mathcal V}_3(\theta_3,\theta_3',\theta_3'',q'',k-q'-q''){\mathcal V}_3(\theta_4,\theta_4',\theta_4'',-q-q'-q'',p+q)\nonumber\\
    &{\mathcal V}_4(\theta_a,\theta_b,\theta_c,\theta_d,p,-q'-q'',k)\biggr) \label{V4}
    \end{align}
    The bosonic and fermionic correlators for the diagram fig \eqref{4ptdiags} are given by
    \begin{align}
    \langle J_0^b(q) J_0^b(q') J_0^b(q'') J_0^b(-q-q'-q'')\rangle &= V^{(1)}_4(q,q',q'',\theta_1,\theta_2,\theta_3,\theta_4)\biggl|_{\theta_1\to 0, \theta_2\to 0,\theta_3\to 0, \theta_4\to 0}\nonumber\\
    \langle J_0^{f}(q) J_0^{f}(q') J_0^{f}(q'') J_0^{f}(-q-q'-q'')\rangle &= \prod_{i=1}^4 \frac{\partial}{\partial\theta_{\alpha i}}\frac{\partial}{\partial\theta^\alpha_i}V^{(1)}_4(q,q',q'',\theta_1,\theta_2,\theta_3,\theta_4)\label{JV}
    \end{align}
    
Although we were able to successfully perform the integrals for the components $p_3,\theta_p $ and $k_3,\theta_k$ in the expression for $V^{(B)}_4$  given by eq.(\ref{V4}) $k_s$ and $p_s$ integrals out be intractable analytically. Due to this difficulty we were not able to obtain a closed form expression for the four point function of the scalar operators$ J_0^b$ and $J_0^{f}$ in eq.(\ref{JV}).

\section{AdS Contact diagrams}\label{Contact}
\subsection{Closed-form}
\begin{align}
&\bar D_{1111}(z, \bar z) =\frac{1}{z-\bar z}[\ln(z\bar z)\ln(\frac{1-z}{1-\bar z})+2\text{Li}_2(z)-2\text{Li}_2(\bar z)] \notag\\
&\bar D_{2222}(z, \bar z) = \frac{12uv}{(z-\bar z)^5}+\frac{1+u+v}{(z-\bar z)^3}[\ln(z\bar z)\ln(\frac{1-z}{1-\bar z})+2\text{Li}_2(z)-2\text{Li}_2(\bar z)]\notag\\&+\frac{6}{(z-\bar z)^4}\bigg((1+u-v)v\ln v + (1+v-u)u \ln u\bigg)+\frac{2}{(z-\bar z)^2}(\ln uv +1)\notag\\
&\bar D_{3333}(u, v) = \notag\\&\bigg(\frac{1680u^2v^2}{(z-\bar z)^9}+\bigg(\frac{240uv}{(z-\bar z)^7}+\frac{24}{(z-\bar z)^5}\bigg)(1+u+v)+\frac{4}{(z-\bar z)^3}\bigg)[\ln(z\bar z)\ln(\frac{1-z}{1-\bar z})+2\text{Li}_2(z)-2\text{Li}_2(\bar z)]\notag\\&+\bigg(\bigg(\frac{840u}{(z-\bar z)^8}+\frac{100}{(z-\bar z)^6}\bigg)v^2(1+u-v)+\frac{480uv}{(z-\bar z)^6}+\frac{12(1+u)+76v}{(z-\bar z)^4})\bigg)\ln v + u \leftrightarrow v\notag\\&+\frac{260uv}{(z-\bar z)^6}+\frac{26}{(z-\bar z)^4}(1+u+v)
\end{align}
\begin{align}
&\bar D(u, v)_{3322} = -\partial_u \bar D_{2222}(u, v)\notag\\
&\bar D(u, v)_{4433} = -\partial_u \bar D_{3333}(u, v)
\end{align}
\subsection{Decomposition in terms of conformal blocks}
The contact diagrams may be written as an expansion in conformal blocks \cite{Hijano:2015zsa}
\begin{align}
D_{\Delta\Delta\Delta'\Delta'}(x_i) = \sum_{m} a^{\Delta\Delta}_m\alpha^{\Delta'\Delta'}_m \mathcal{W}_{\Delta_m,0}(x_i) + \sum_{n}a^{\Delta\Delta}_n\alpha^{\Delta'\Delta'}_n \mathcal{W}_{\Delta_n,0}(x_i)
\end{align}
\begin{align}
D_{\Delta\Delta\Delta\Delta}(x_i) = \sum_{n} 2a^{\Delta\Delta}_n(\sum_{m\neq n}\frac{a^{\Delta\Delta}_m}{m^2_n-m^2_m}) \mathcal{W}_{\Delta_n,0}(x_i) + \sum_{n}(a^{\Delta\Delta}_n)^2 \frac{\partial}{\partial m^2_n}\mathcal{W}_{\Delta_n,0}(x_i)
\end{align}
where $\mathcal{W}_{\Delta, 0} = \beta_{\Delta 34}\beta_{\Delta 12}W_{\Delta, 0}$. For $\Delta_i = \Delta$
\begin{align}
D_{\Delta\Delta\Delta\Delta}(x_i) &= \sum_{n} 2a^{\Delta\Delta}_n\eta^{\Delta\Delta}_n \mathcal{W}_{\Delta_n,0}(x_i) + \sum_{n}(a^{\Delta\Delta}_n)^2 \frac{\partial}{\partial m^2_n}\mathcal{W}_{\Delta_n,0}(x_i)
\\\notag&=\sum_n[(2a^{\Delta\Delta}_n\eta^{\Delta\Delta}_n+(a^{\Delta\Delta}_n)^2)\beta^2_{\Delta_n\Delta\Delta}+\frac{\partial}{\partial m^2_n}\beta^2_{\Delta_n\Delta\Delta}]W_{\Delta_n, 0}(x_i)\notag\\&+\sum_{n}(a^{\Delta\Delta}_n)^2 \beta^2_{\Delta_n\Delta\Delta}\frac{\partial}{\partial m^2_n}W_{\Delta_n,0}(x_i)
\end{align}
with
\begin{align}
&\eta^{\Delta\Delta}_n = \sum_{m \neq n}\frac{a^{\Delta\Delta}_m}{m^2_n-m^2_m}\\
&\beta_{\Delta 34} \equiv {\Gamma\left({\Delta+\Delta_{34}\over 2}\right)\Gamma\left({\Delta-\Delta_{34}\over 2}\right) \over 2\Gamma (\Delta)}
\\\notag
&m_{\Delta_k}^2 = \Delta_k(\Delta_k-d)\\
&a^{12}_{m} = \frac{(-1)^m}{\beta_{\Delta_m12}m!}\frac{(\Delta_1)_m(\Delta_2)_m}{(\Delta_1+\Delta_2+m-d/2)_m}
\end{align}
with the anomalous dimension being proportional to the coefficient of the third term which involves derivative of the conformal block.
Writing the above interms of the $\bar D$ functions
\begin{align}
\bar D_{\Delta\Delta\Delta\Delta}(u, v) &=\frac{1}{u^{\Delta}}[\sum_n[(2a^{\Delta\Delta}_n\eta^{\Delta\Delta}_n)\beta^2_{\Delta_n\Delta\Delta}+(a^{\Delta\Delta}_n)^2\frac{\partial}{\partial m^2_n}\beta^2_{\Delta_n\Delta\Delta}]G_{\Delta_n, 0}(u, v)\notag\\&+\sum_{n}(a^{\Delta\Delta}_n)^2 \beta^2_{\Delta_n\Delta\Delta}\frac{\partial}{\partial m^2_n}G_{\Delta_n,0}(u, v)]
\end{align}
We will re-label
\begin{align}
&P^{(\Delta)}_1(n, 0)=(2a^{\Delta\Delta}_n\eta^{\Delta\Delta}_n+(a^{\Delta\Delta}_n)^2)\beta^2_{\Delta_n\Delta\Delta}+\frac{\partial}{\partial m^2_n}\beta^2_{\Delta_n\Delta\Delta}\notag\\
&P^{(\Delta)}_0(n, 0)\gamma^{(\Delta)}_1(n, 0) = 2(a^{\Delta\Delta}_n)^2 \beta^2_{\Delta_n\Delta\Delta}
\end{align}
so that 
\begin{align}
\bar D_{\Delta\Delta\Delta\Delta}(u, v) &=\frac{2\Gamma(\Delta)^4}{\Gamma(2\Delta-d/2)}\frac{1}{u^{\Delta}}\sum_{n}[P^{(\Delta)}_1(n, 0)G_{\Delta_n, 0}(u, v)\notag\\&+\frac{1}{2}P^{(\Delta)}_0(n, 0)\gamma^{(\Delta)}_1(n, 0)\frac{\partial}{\partial m^2_n}G_{\Delta_n,0}(u, v)]
\end{align}
satisfying \cite{Heemskerk:2009pn}
\begin{align}
P^{(\Delta)}_1(n, 0) = \frac{1}{2}\partial_n(P^{(\Delta)}_0(n, 0)\gamma^{(\Delta)}_{1}(n, 0))
\end{align}
Similarly, for (141)
\begin{align}
\bar D_{\Delta+1\Delta+1\Delta\Delta}(u, v) =&\frac{2\Gamma(\Delta)^2\Gamma(\Delta+1)^2}{\Gamma(2\Delta+1-d/2)} \frac{1}{u^\Delta}\notag\\&[\sum_{m} \bar P^{(\Delta)}_1(n, 0) G_{\Delta_m,0}(u, v) +\frac{1}{2}\bar P^{(\Delta)}_0(n, 0)\bar\gamma^{(\Delta)}_1(n, 0)\frac{\partial}{\partial m^2_n}G_{\Delta_n,0}(u, v)]\notag\\&+ \beta^2_{2\Delta}a^{\Delta\Delta}_0\eta^{\Delta+1\Delta+1}_0G_{2\Delta,0}(u, v)
\end{align}
\begin{align}
&\bar P^{(\Delta)}_1(n, 0)=(a^{\Delta+1\Delta+1}_n\eta^{\Delta\Delta}_n+a^{\Delta\Delta}_n\eta^{\Delta+1\Delta+1}_n)\beta^2_{\Delta_n\Delta\Delta}+a^{\Delta\Delta}_na^{\Delta+1\Delta+1}_n\frac{\partial}{\partial m^2_n}\beta^2_{\Delta_n\Delta\Delta}\notag\\
&\bar P^{(\Delta)}_0(n, 0)\bar\gamma^{(\Delta)}_1(n, 0) = 2a^{\Delta\Delta}_na^{\Delta+1\Delta+1}_n \beta^2_{\Delta_n\Delta\Delta}
\end{align}
\subsubsection{Examples}
\begin{align}
&\bar D_{1111}(u, v) =\frac{2}{\pi^{1/2}u}\sum_{n}[P^{(1)}_1(n, 0)G_{2+2n, 0}(u, v)+\frac{1}{2}P^{(1)}_0(n, 0)\gamma^{(1)}_1(n, 0)\frac{\partial_nG_{2+2n,0}(u, v)}{8n+2}]\notag\\
&\bar D_{2222}(u, v) =\frac{8}{3\pi^{1/2}u^2}\sum_{n}[P^{(2)}_1(n, 0)G_{4+2n, 0}(u, v)+\frac{1}{2}P^{(2)}_0(n, 0)\gamma^{(2)}_1(n, 0)\frac{\partial_nG_{4+2n,0}(u, v)}{8n+10}]\notag\\
&\bar D_{3333}(u, v) =\frac{256}{105\pi^{1/2}u^3}\sum_{n}[P^{(3)}_1(n, 0)G_{6+2n, 0}(u, v)+\frac{1}{2}P^{(3)}_0(n, 0)\gamma^{(3)}_1(n, 0)\frac{\partial_nG_{6+2n,0}(u, v)}{8n+18}]\label{ContactCD1}
\end{align}

\begin{align}
&\bar D_{3322}(u, v) =\frac{64}{15\pi^{1/2}u^3}[\sum_{m} \bar P^{(3)}_1(m, 0) G_{6+2m,0}(u, v) +\frac{1}{2}\bar P^{(3)}_0(m, 0)\bar\gamma^{(3)}_1(m, 0)\frac{\partial}{\partial m^2_n}G_{6+2m,0}(u, v)\notag\\&+ \beta^2_{4}a^{22}_0\eta^{33}_0G_{4,0}(u, v)]\notag\\
&\bar D_{4433}(u, v) =\frac{1024}{105 \sqrt{\pi }u^4}[\sum_{m} \bar P^{(4)}_1(m, 0) G_{8+2m,0}(u, v) +\frac{1}{2}\bar P^{(4)}_0(m, 0)\bar\gamma^{(4)}_1(m, 0)\frac{\partial}{\partial m^2_n}G_{8+2m,0}(u, v)\notag\\&+ \beta^2_{6}a^{33}_0\eta^{44}_0G_{6,0}(u, v)]
\label{ContactCD2}
\end{align}
\textbf{Contact terms for bosonic correlator}
\begin{align}
&G^{AdS}_{\phi^4} = \bar D_{1111} (u, v)\notag\\
&G^{AdS}_{(\partial\phi)^4} = (1+u+v)\bar D_{2222} (u, v) \notag\\
&G^{AdS}_{\phi^2(\partial^3\phi)^2} =2(u^2 \bar D_{3322}(u, v) +v^2 \bar D_{3322}(v, u)+ \frac{1}{v^3}\bar D_{3322}(1/v, u/v))\label{ContactBB}
\end{align}
\textbf{Contact terms for fermionic correlator}
\begin{align}
&G^{AdS}_{\phi^4} = \bar D_{2222} (u, v)\notag\\
&G^{AdS}_{(\partial\phi)^4} = (1+u+v)\bar D_{3333} (u, v) \notag\\
&G^{AdS}_{\phi^2(\partial^3\phi)^2} =2(u^2 \bar D_{4433}(u, v) +v^2 \bar D_{4433}(v, u)+  \frac{1}{v^3}\bar D_{4433}(1/v, u/v))\label{ContactFF}
\end{align}
\section{Computing the coefficients}\label{apf}

\subsection{Bosonic case}
 We have derived similar relations for $\Delta \geq 4$ i.e. $n \geq 1$
\begin{align}
[\tilde C^{2(BBB)}_{0, susy}]_{[O, O]_{n, 0}} &=  \frac{1+\tilde{\lambda}^4}{\tilde N (1+\tilde{\lambda}^2)^2}[\tilde C^{2}_{0, fb}]_{[O, O]_{n, 0}}+ c_1 [\frac{2}{\pi^{1/2}}P^{(1)}_1(n, 0)] +c_2[\frac{8}{3\pi^{1/2}}P^{(2)}_1(n-1, 0)]\notag\\&+c_3\frac{64}{15\pi^{1/2}} [\bar P^{(3)}_{1}(n-2, 0)\Theta(n-1) + \beta^2_{4}a^{22}_0\eta^{33}_0\delta_{n-1, 0} ]
\end{align}
Let us take two arbitrary values of $n$ to solve for $c_2$ and $c_3$. Let them be $i, j$. So, the solution is, therefore,
\begin{align}
&c_2 = \frac{E[j] (A[i]+B[i]-C[i])-E[i] (A[j]+B[j]-C[j])}{D[j] E[i]-D[i] E[j]}\\
&c_3=\frac{D[j] (A[i]+B[i]-C[i])-D[i] (A[j]+B[j]-C[j])}{D[j] E[i]-D[i] E[j]}
\end{align}
where
\begin{align}
&A[n] = \frac{1+\tilde{\lambda}^4}{\tilde N (1+\tilde{\lambda}^2)^2}[\tilde C^{2}_{0, fb}]_{[O, O]_{n, 0}}\notag\\
&B[n] = c_1 [\frac{2}{\pi^{1/2}}P^{(1)}_1(n, 0)]\notag\\
&C[n]=[\tilde C^{2(BBB)}_{0, susy}]_{[O, O]_{n, 0}}\notag\\
&D[n] = \frac{8}{3\pi^{1/2}}P^{(2)}_1(n-1, 0)\notag\\
&E[n] = \frac{64}{15\pi^{1/2}} [\bar P^{(3)}_{1}(n-2, 0)\Theta(n-1) + \beta^2_{4}a^{22}_0\eta^{33}_0\delta_{n-1, 0} ]
\end{align}
When $i = 1, j = 2$, we get

\begin{align}
&[\tilde C^{2(BBB)}_{0, susy}]_{[O, O]_{1, 0}} =  \frac{1+\tilde{\lambda}^4}{\tilde N (1+\tilde{\lambda}^2)^2}[\tilde C^{2}_{0, fb}]_{[O, O]_{1, 0}}+ c_1 [-0.0113728] +c_2[1.14329]+c_3[0.0891028]\notag\\
&[\tilde C^{2(BBB)}_{0, susy}]_{[O, O]_{2, 0}} =  \frac{1+\tilde{\lambda}^4}{\tilde N (1+\tilde{\lambda}^2)^2}[\tilde C^{2}_{0, fb}]_{[O, O]_{2, 0}}+ c_1 [5.78456*10^-6] +c_2[0.0151638]+c_3[9.58344]
\end{align}
so that
\begin{align}
&c_2 = -0.874774~ \frac{1+\tilde{\lambda}^4}{\tilde N (1+\tilde{\lambda}^2)^2}[\tilde C^{2}_{0, fb}]_{[O, O]_{1, 0}}+0.00793315~ \frac{1+\tilde{\lambda}^4}{\tilde N (1+\tilde{\lambda}^2)^2}[\tilde C^{2}_{0, fb}]_{[O, O]_{2, 0}}\notag\\&+0.874774 [\tilde C^{2(BBB)}_{0, susy}]_{[O, O]_{1, 0}}-0.00793315~ [\tilde C^{2(BBB)}_{0, susy}]_{[O, O]_{2, 0}}+0.00994867 c_1\\
&c_3=0.00138415~ \frac{1+\tilde{\lambda}^4}{\tilde N (1+\tilde{\lambda}^2)^2}[\tilde C^{2}_{0, fb}]_{[O, O]_{1, 0}}-0.104359 \frac{1+\tilde{\lambda}^4}{\tilde N (1+\tilde{\lambda}^2)^2}[\tilde C^{2}_{0, fb}]_{[O, O]_{2, 0}}\notag\\&-0.00138415 [\tilde C^{2(BBB)}_{0, susy}]_{[O, O]_{1, 0}}+0.104359~ [\tilde C^{2(BBB)}_{0, susy}]_{[O, O]_{2, 0}}-0.0000163453~ c_1
\end{align}
\subsection{Fermionic case}
We compute the OPE coefficient relations for scalar double-trace operators with $\Delta \geq 4$ using the methodology of the previous subsection.

\begin{align}
[\tilde C^{2(FFF)}_{0, susy}]_{[O, O]_{n, 0}} &=  \frac{1+\tilde{\lambda}^4}{\tilde N (1+\tilde{\lambda}^2)^2}[\tilde C^{2}_{0, ff}]_{[O, O]_{n, 0}}+ \bar c_1[\frac{8}{3\pi^{1/2}}P^{(2)}_1(n, 0)]+\bar c_2[\frac{256}{105\pi^{1/2}}\Theta(n)P^{(3)}_1(n-1, 0)]\notag\\&+\bar c_3\frac{1024}{105 \sqrt{\pi }} [\bar P^{(4)}_1(n-2, 0) \Theta(n-1) + \beta^2_{6}a^{33}_0\eta^{44}_0 \delta_{n-1, 0} ]
\end{align}
\bibliography{mc.bib}

\end{document}